\documentclass[11pt,a4paper]{llncs}
\usepackage[english]{babel}
\usepackage{german}
\usepackage{amssymb}
\usepackage{amsfonts}
\usepackage{amsmath}
\usepackage[latin1]{inputenc}
\usepackage{fullpage}
\usepackage{graphicx}
\usepackage{subfigure}
\usepackage{algorithm}
\usepackage[noend]{algorithmic}
\usepackage{amsmath,amssymb,cite}
\usepackage{lineno,footmisc,marvosym}
\usepackage{wasysym,vmargin,stackrel}

\newtheorem{observation}{Observation}
\setmarginsrb{3.5cm}{2.25cm}{3.5cm}{3.05cm}{0.3cm}{0.3cm}{-0.3cm}{1.0cm}
   \renewenvironment{thebibliography}[1]{
 \begin{oldthebibliography}{#1}
 \setlength{\parskip}{0.0ex} \setlength{\itemsep}{1ex}}  {\end{oldthebibliography}}

\begin{document}

\title{Strong Bounds for Evolution in Networks\thanks{This work was partially supported by the EPSRC Grants~EP/P020372/1 and~EP/P02002X/1.} 
\thanks{A preliminary conference version of this work appeared in the \emph{Proceedings of the 40th International 
Colloquium on Automata, Languages and Programming (ICALP)}, Riga, Latvia, pages 675--686, 2013.}}

\author{George B. Mertzios\inst{1} \and Paul G. Spirakis\inst{2}}
\institute{School of Engineering and Computing Sciences, Durham University, UK. 
Email: \texttt{george.mertzios@durham.ac.uk} 
\and
Department of Computer Science, University of Liverpool, UK, and University of Patras, Greece.
Email: \texttt{p.spirakis@liverpool.ac.uk}\vspace{-0.2cm}}
\date{\vspace{-5cm}}
\maketitle

\begin{abstract}
This work studies the generalized Moran process, as introduced by Lieberman et al. [Nature, 433:312-316, 2005]. We introduce the parameterized notions of \emph{selective amplifiers} and \emph{selective suppressors} of evolution, i.e. of networks (graphs) with many ``strong starts'' and many ``weak starts'' for the mutant, respectively. We first prove the existence of \emph{strong} selective amplifiers and of (quite) strong selective suppressors. Furthermore we provide strong upper bounds and almost tight lower bounds (by proving the ``Thermal Theorem'') for the traditional notion of fixation probability of Lieberman et al., i.e. assuming a random initial placement of the mutant.\newline

\noindent \textbf{Keywords:} Evolutionary dynamics, undirected graphs,
fixation probability, lower bound, Markov chain.
\end{abstract}

\section{Introduction}

\label{intro-sec}

Population and evolutionary dynamics have been extensively studied~\cite%
{Traulsen09,Taylor06,Taylor04,Ohtsuki06,Karlin75,Broom09,Antal06}, mainly on
the assumption that the evolving population is homogeneous, i.e.~it has no
spatial structure. One of the main models in this area is the Moran Process~%
\cite{Moran58}, where the initial population contains a single \emph{mutant}
with fitness $r>0$, with all other individuals having fitness $1$. At every
step of this process, an individual is chosen for reproduction with
probability proportional to its fitness. This individual then replaces a
second individual, which is chosen uniformly at random, with a copy of
itself. Such dynamics as the above have been extensively studied also in the
context of strategic interaction in evolutionary game theory~\cite%
{Gintis00,Sandholm11,Hofbauer98,Kandori93,Imhof05}.

In a recent article, Lieberman, Hauert, and Nowak~\cite{Nowak05} (see also~%
\cite{nowak06-book}) introduced a generalization of the Moran process, where
the individuals of the population are placed on the vertices of a connected
graph (which is, in general, directed) such that the edges of the graph
determine competitive interaction. In the generalized Moran process, the
initial population again consists of a single mutant of fitness $r$, placed
on a vertex that is chosen uniformly at random, with each other vertex
occupied by a non-mutant of fitness $1$. An individual is chosen for
reproduction exactly as in the standard Moran process, but now the second
individual to be replaced is chosen among its neighbors in the graph
uniformly at random (or according to some weights of the edges)~\cite%
{Nowak05,nowak06-book}. If the underlying graph is the complete graph, then
this process becomes the standard Moran process on a homogeneous population~%
\cite{Nowak05,nowak06-book}. Several similar models describing infections
and particle interactions have been also studied in the past, including
the~SIR and~SIS epidemics~\cite[Chapter~21]{EasleyK10}, the voter and
antivoter models and the exclusion process~\cite%
{Liggett85,Aldous-online-book,Durrett88}. However such models do not
consider the issue of different fitness of the individuals.

The central question that emerges in the generalized Moran process is how
the population structure affects evolutionary dynamics~\cite%
{Nowak05,nowak06-book}. In the present work we consider the generalized
Moran process on arbitrary finite, undirected, and connected graphs. On such
graphs, the generalized Moran process terminates almost surely, reaching
either \emph{fixation} of the graph (all vertices are occupied by copies of
the mutant) or \emph{extinction} of the mutants (no copy of the mutant
remains). The \emph{fixation probability} of a graph $G$ for a mutant of
fitness $r$, is the probability that eventually fixation is reached when the
mutant is initially placed at a random vertex of $G$, and is denoted by $%
f_{r}(G)$. The fixation probability can, in principle, be determined using
standard Markov Chain techniques. But doing so for a general graph on $n$
vertices requires solving a linear system of~ $2^{n}$ linear equations. Such
a task is not computationally feasible, even numerically. As a result of
this, most previous work on computing fixation probabilities in the
generalized Moran process was either restricted to graphs of small size~\cite%
{Broom09} or to graph classes which have a high degree of symmetry, reducing
thus the size of the corresponding linear system (e.g.~paths, cycles, stars,
and cliques~\cite{Broom08,Broom-two-results,Broom10}). Experimental results
on the fixation probability of random graphs derived from grids can be found
in~\cite{RS2008:Smallworlds}. Moreover, layered directed graphs with extreme behavior 
due to the existence of ``positive feedback loops'', such as superstars, megastars, funnels, and metafunnels, 
have been studied in~\cite{Nowak05,DiazGoldMerRicSerSpir_Superstar_2013,GalanisGGLR17}.

A recent result~\cite{DiazGMRSS12-algorithmica14} shows how to
construct fully polynomial randomized approximation schemes (FPRAS) for the
probability of reaching fixation (when $r\geq 1$) or extinction (for all $%
r>0 $). The result of~\cite{DiazGMRSS12-algorithmica14} uses a Monte Carlo estimator,
i.e.~it runs the generalized Moran process several times\footnote{%
For approximating the probability to reach fixation (resp.~extinction), one
needs a number of runs which is about the inverse of the best known lower
(resp.~upper) bound of the fixation probability.}, while each run terminates
in polynomial time with high probability~\cite{DiazGMRSS12-algorithmica14}. Note that
improved lower and upper bounds on the fixation probability immediately lead
to a better estimator here. Until now, the only known general bounds for the
fixation probability on connected undirected graphs, are that $f_{r}(G)\geq 
\frac{1}{n}$ and $f_{r}(G)\leq 1-\frac{1}{n+r}$.

Lieberman et al.~\cite{Nowak05,nowak06-book} proved the \emph{Isothermal Theorem}, 
stating that (in the case of undirected graphs) the fixation probability of a regular graph 
(i.e.~of a graph with overall the same vertex degree) is equal to that of the complete graph 
(i.e.~the homogeneous population of the standard Moran process), which equals 
to $(1-\frac{1}{r}) / (1-\frac{1}{r^{n}})$, where $n$ is the size of the population. 
Intuitively, in the Isothermal Theorem, every vertex of the graph has a \emph{temperature} which determines
how often this vertex is being replaced by other individuals during the
generalized Moran process. 
The complete graph (or equivalently, any regular graph) serves as a benchmark for
measuring the fixation probability of an arbitrary graph $G$: if $f_{r}(G)$
is larger (resp.~smaller) than that of the complete graph then $G$ is called
an \emph{amplifier} (resp.~a \emph{suppressor})~\cite{Nowak05,nowak06-book}.

\vspace{0.2cm}

\noindent \textbf{Our contribution.} The structure of the graph, on which
the population resides, plays a crucial role in the course of evolutionary
dynamics. Human societies or social networks are never homogeneous, while
certain individuals in central positions may be more influential than others~%
\cite{nowak06-book}. Motivated by this, we introduce in this paper a new
notion of measuring the success of an advantageous mutant in a structured
population, by counting the number of initial placements of the mutant in a
graph that guarantee fixation of the graph with large probability. This
provides a refinement of the notion of fixation probability. 
Specifically, we do not any more consider the fixation probability
as the probability of reaching fixation when the mutant is placed at a
random vertex, but we rather consider the probability $f_{r}(v)$ of reaching
fixation when a mutant with fitness $r>1$ is introduced at a specific vertex 
$v$ of the graph; $f_{r}(v)$ is termed the \emph{fixation probability of
vertex $v$}. Using this notion, the fixation probability $f_{r}(G)$ of a
graph $G=(V,E)$ with $n$ vertices is $f_{r}(G)=\frac{1}{n}\sum_{v\in
V}f_{r}(v)$.

We aim in finding graphs that have many ``strong starts'' (or many ``weak
starts'') of the mutant. Thus we introduce the notions of $(h(n),g(n))$\emph{%
-selective amplifiers} (resp.~$(h(n),g(n))$\emph{-selective suppressors}),
which include those graphs with $n$ vertices for which there exist at least $%
h(n)$ vertices $v$ with $f_{r}(v)\geq 1-\frac{c(r)}{g(n)}$ (resp.~$%
f_{r}(v)\leq \frac{c(r)}{g(n)}$) for an appropriate function $c(r)$ of $r$.
We contrast this new notion of $(h(n),g(n))$-selective amplifiers
(resp.~suppressors) with the notion of $g(n)$\emph{-universal amplifiers}
(resp.~\emph{suppressors}) which include those graphs $G$ with $n$ vertices
for which $f_{r}(G)\geq 1-\frac{c(r)}{g(n)}$ (resp.~$f_{r}(G)\leq \frac{c(r)%
}{g(n)}$) for an appropriate function $c(r)$ of $r$. For a detailed
presentation and a rigorous definition of these notions we refer to Section~%
\ref{preliminaries-sec}.

Using these new notions, we prove that there exist strong selective amplifiers, namely $(\Theta (n),n)$-selective amplifiers 
(called the \emph{urchin graphs}). Furthermore we 
prove that there exist also quite strong selective suppressors, namely 
$(\frac{n}{\phi (n)+1},\frac{n}{\phi (n)})$-selective suppressors (called the $\phi(n)$\emph{-urchin graphs}) 
for \emph{any} function $\phi (n)=\omega (1)$ with $\phi (n)\leq \sqrt{n}$.

Regarding the traditional measure of the fixation probability $f_{r}(G)$ of undirected graphs $G$, 
we provide upper and lower bounds that are much stronger than the bounds $\frac{1}{n}$ and $1-\frac{1}{n+r}$ 
that were known so far~\cite{DiazGMRSS12-algorithmica14}. 
More specifically, first of all we
demonstrate the nonexistence of ``strong'' universal amplifiers by showing
that for any graph $G$ with $n$ vertices, the fixation probability $f_{r}(G)$ is 
strictly less than $1-\frac{c(r)}{g(n)}$, for any $g(n)=\omega(n^{\frac{3}{4}})$. 
This is in a wide contrast with what happens in directed
graphs, as Lieberman et al.~\cite{Nowak05} provided directed graphs with
arbitrarily large fixation probability (see also~\cite{nowak06-book}).
Motivated by our work, very recently Giakkoupis~\cite{Giakkoupis16} and Goldberg et al.~\cite{GoldbergLLMPP16} slightly improved 
our estimate of the function $g(n)$ to $\omega(n^{\frac{1}{3}} \log^{\frac{4}{3}} n)$~\cite{Giakkoupis16} and to $\omega(n^{\frac{1}{3}})$~\cite{GoldbergLLMPP16}, respectively.

On the other hand, we provide our lower bound in the \emph{Thermal Theorem},
which states that for any vertex $v$ of an arbitrary undirected graph $G$,
the fixation probability $f_{r}(v)$ of $v$ is at least $(r-1)/(r+\frac{\deg v%
}{\deg _{\min }})$ for any $r>1$, where $\deg v$ is the degree of $v$ in $G$
(i.e.~the number of its neighbors) and $\deg _{\min }$ (resp.~$\deg _{\max }$%
) is the minimum (resp.~maximum) degree in $G$. This result extends the
Isothermal Theorem for regular graphs~\cite{Nowak05}. In particular, we
consider here a different notion of \emph{temperature} for a vertex than~%
\cite{Nowak05}: the temperature of vertex $v$ is $\frac{1}{\deg v}$. As it
turns out, a ``hot'' vertex (i.e.~with high temperature) affects more often
its neighbors than a ``cold'' vertex (with low temperature). The Thermal
Theorem, which takes into account the vertex $v$ on which the mutant is
introduced, provides immediately our lower bound $(r-1)/(r+%
\frac{\deg _{\max }}{\deg _{\min }})$ for the fixation probability $f_{r}(G)$
of any undirected graph $G$. The latter lower bound is almost tight, as it
implies that $f_{r}(G)\geq \frac{r-1}{r+1}$ for a regular graph $G$, while
the Isothermal Theorem implies that the fixation probability of a regular
graph $G$ tends to $\frac{r-1}{r}$ as the size of $G$ increases. Note that
our new upper/lower bounds for the fixation probability lead to better time
complexity of the FPRAS proposed in~\cite{DiazGMRSS12-algorithmica14}, as the Monte
Carlo technique proposed in~\cite{DiazGMRSS12-algorithmica14} now needs to simulate
the Moran process a less number of times (to estimate fixation or
extinction).

Our techniques are original and of a constructive combinatorics flavor. For
the class of strong selective amplifiers (the urchin graphs)
we introduce a novel decomposition of the Markov chain $\mathcal{M}$ of the
generalized Moran process into $n-1$ smaller chains $\mathcal{M}_{1},%
\mathcal{M}_{2},\ldots ,\mathcal{M}_{n-1}$, and then we decompose each $%
\mathcal{M}_{k}$ into two even smaller chains $\mathcal{M}_{k}^{1},\mathcal{M%
}_{k}^{2}$. Then we exploit a new way of composing these smaller chains (and
returning to the original one) that is carefully done to maintain the needed
domination properties. For the proof of the lower bound in the Thermal
Theorem, we first introduce a new and simpler weighted process that bounds
fixation probability from below (the generalized Moran process is a special
case of this new process). Then we add appropriate dummy states to its
(exponentially large) Markov chain, and finally we iteratively modify the
resulting chain by maintaining the needed monotonicity properties.
Eventually this results to the desired lower bound of the Thermal Theorem.
Finally, our proof for the non-existence of strong universal amplifiers is
done by contradiction, partitioning appropriately the vertex set of the
graph and discovering an appropriate independent set that leads to the
contradiction.

The paper is organized as follows. Preliminaries and notation are given in
Section~\ref{preliminaries-sec}. Furthermore we present our results on
amplifiers and suppressors in Sections~\ref{amplifiers-sec} and~\ref%
{suppressors-sec}, respectively.

\section{Preliminaries\label{preliminaries-sec}}

Throughout the paper we consider only finite, connected, undirected graphs $%
G=(V,E)$. Our results apply to connected graphs as, otherwise, the fixation
probability is necessarily zero. The edge $e\in E$ between two vertices $%
u,v\in V$ is denoted by $e=uv$. For a vertex subset $X\subseteq V$, we write 
$X+y$ and $X-y$ for $X\cup \{y\}$ and $X\setminus \{y\}$, respectively.
Furthermore, throughout $r$ denotes the fitness of the mutant, while the
value $r$ is considered to be independent of the size $n$ of the network,
i.e.~we assume that $r$ is constant. For simplicity of presentation, we call
a vertex $v$ ``infected'' if a copy of the mutant is placed on $v$. For
every vertex subset $S\subseteq V$ we denote by $f_{r}(S)$ the fixation
probability of the set $S$, i.e.~the probability that, starting with exactly 
$|S|$ copies of the mutant placed on the vertices of $S$, the generalized
Moran process will eventually reach fixation. By the definition of the
generalized Moran process $f_{r}(\emptyset )=0$ and $f_{r}(V)=1$, 
while for $S\notin \{\emptyset ,V\}$,
\begin{equation}
f_{r}(S)=\frac{\sum_{xy\in E,x\in S,y\notin S}\left( \frac{r}{\deg x}%
f_{r}(S+y)+\frac{1}{\deg y}f_{r}(S-x)\right) }{\sum_{xy\in E,x\in S,y\notin
S}\left( \frac{r}{\deg x}+\frac{1}{\deg y}\right) }
\label{generalized-Moran-exact-fixation-eq}
\end{equation}%
In the next definition we introduce the notions of \emph{universal} and 
\emph{selective} amplifiers.

\begin{definition}
\label{amplifiers-def}Let $\mathcal{G}$ be an infinite class of undirected
graphs. If there exists an $n_{0}\in \mathbb{N}$, an $r_{0}\geq 1$, and some
function $c(r)$, such that for every graph $G\in \mathcal{G}$ with $n\geq
n_{0}$ vertices and for every $r>r_{0}$:

\begin{itemize}
\item $f_{r}(G)\geq 1-\frac{c(r)}{g(n)}$, then $\mathcal{G}$ is a class of $%
g(n)$\emph{-universal amplifiers},

\item there exists a subset $S$ of at least $h(n)$ vertices of $G$, such
that $f_{r}(v)\geq 1-\frac{c(r)}{g(n)}$ for every vertex $v\in S$, then $%
\mathcal{G}$ is a class of $(h(n),g(n))$\emph{-selective amplifiers}.
\end{itemize}

Moreover, $\mathcal{G}$ is a class of \emph{strong universal} (resp.~\emph{%
strong selective}) amplifiers if $\mathcal{G}$ is a class of $n$\emph{%
-universal} (resp.~$(\Theta (n),n)$\emph{-selective}) amplifiers.
\end{definition}

Similarly to Definition~\ref{amplifiers-def}, we introduce the notions of 
\emph{universal} and \emph{selective} suppressors.

\begin{definition}
\label{suppressors-def}Let $\mathcal{G}$ be an infinite class of undirected
graphs. If there exist functions $c(r)$ and $n_{0}(r)$, such that for every $%
r>1$ and for every graph $G\in \mathcal{G}$ with $n\geq n_{0}(r)$ vertices:

\begin{itemize}
\item $f_{r}(G)\leq \frac{c(r)}{g(n)}$, then $\mathcal{G}$ is a class of $%
g(n)$\emph{-universal suppressors},

\item there exists a subset $S$ of at least $h(n)$ vertices of $G$, such
that $f_{r}(v)\leq \frac{c(r)}{g(n)}$ for every vertex $v\in S$, then $%
\mathcal{G}$ is a class of $(h(n),g(n))$\emph{-selective suppressors}.
\end{itemize}

Moreover, $\mathcal{G}$ is a class of \emph{strong universal} (resp.~\emph{%
strong selective}) suppressors if $\mathcal{G}$ is a class of $n$\emph{%
-universal} (resp.~$(\Theta (n),n)$\emph{-selective}) suppressors.
\end{definition}

Note that $n_{0}=n_{0}(r)$ in Definition~\ref{suppressors-def}, while in
Definition~\ref{amplifiers-def} $n_{0}$ is not a function of $r$. The reason
for this is that, since we consider the fitness value $r$ to be constant,
the size $n$ of $G$ needs to be sufficiently large with respect to $r$ in
order for $G$ to act as a suppressor. Indeed, if we let $r$ grow
arbitrarily, e.g.~if $r=n^{2}$, then for \emph{any} graph $G$ with $n$
vertices the fixation probability $f_{r}(v)$ tends to $1$ as $n$ grows. The
next lemma follows by Definitions~\ref{amplifiers-def} and~\ref%
{suppressors-def}.

\begin{lemma}
\label{universal-selective-lem}If $\mathcal{G}$ is a class of $g(n)$%
-universal amplifiers (resp.~suppressors), then $\mathcal{G}$ is a class of $%
(n-o(n),g(n))$-selective amplifiers (resp.~suppressors).
\end{lemma}

\begin{proof}
Suppose that $\mathcal{G}$ is a class of $g(n)$-universal amplifiers. That
is, for every ${r>r}_{0}$ and for every graph ${G=(V,E)\in \mathcal{G}}$
with~${n\geq n_{0}}$ vertices, the fixation probability of $G$ is $%
f_{r}(G)\geq 1-\frac{c(r)}{g(n)}$, where $c(r)$ is some function that
depends only on $r$. Let $S\subseteq V$ be the subset of vertices such that $%
f_{r}(v)\geq 1-\frac{c^{\prime }(r)}{g(n)}$ for some function $c^{\prime
}(r) $ that depends only on $r$. Then there exists an appropriate function $%
\phi (n,r)=\omega (1)$, i.e.~$\underset{n\rightarrow \infty }{\lim }\phi
(n,r)=\infty $, such that ${f_{r}(v)\leq 1-\frac{\phi (n,r)}{g(n)}}$ for
every $v\in V\setminus S$. Thus the fixation probability of~$G$~is 
\begin{equation}
f_{r}(G)\leq \frac{|S|\cdot 1+(n-|S|)\cdot (1-\frac{\phi (n,r)}{g(n)})}{n}=1-%
\frac{(n-|S|)}{n}\cdot \frac{\phi (n,r)}{g(n)}  \label{eq-1}
\end{equation}%
Now, since $f_{r}(G)\geq 1-\frac{c(r)}{g(n)}$, it follows by (\ref{eq-1})
that $(n-|S|)\leq n\frac{c(r)}{\phi (n,r)}$, and thus $|S|\geq n(1-\frac{c(r)%
}{\phi (n,r)})=n-o(n)$, since $\phi (n,r)=\omega (1)$. Thus it follows
by definition of the set $S$ that $\mathcal{G}$ is a class of ${(n-o(n),g(n))}$-selective amplifiers.

Suppose now that $\mathcal{G}$ is a class of $g(n)$-universal suppressors.
That is, for every $r>1$ and for every graph ${G=(V,E)\in \mathcal{G}}$ with~%
$n\geq n_{0}(r)$ vertices, the fixation probability of $G$ is $f_{r}(G)\leq 
\frac{c(r)}{g(n)}$, where $c(r)$ is some function that depends only on $r$.
Let $S\subseteq V$ be the subset of vertices such that $f_{r}(v)\leq \frac{%
c^{\prime }(r)}{g(n)}$ for some function $c^{\prime }(r)$ that depends only
on $r$. Then there exists an appropriate function $\phi (n,r)=\omega (1)$,
i.e.~$\underset{n\rightarrow \infty }{\lim }\phi (n,r)=\infty $, such that ${%
f_{r}(v)\geq \frac{\phi (n,r)}{g(n)}}$ for every $v\in V\setminus S$. Thus
the fixation probability of~$G$~is%
\begin{equation}
f_{r}(G)\geq \frac{|S|\cdot 0+(n-|S|)\cdot \frac{\phi (n,r)}{g(n)}}{n}=\frac{%
(n-|S|)}{n}\cdot \frac{\phi (n,r)}{g(n)}  \label{eq-2}
\end{equation}%
Now, since $f_{r}(G)\leq \frac{c(r)}{g(n)}$, it follows by (\ref{eq-2}) that 
$(n-|S|)\leq n\frac{c(r)}{\phi (n,r)}$, and thus $|S|\geq n(1-\frac{c(r)}{%
\phi (n,r)})=n-o(n)$, since $\phi (n,r)=\omega (1)$. Thus it follows by
definition of the set $S$ that $\mathcal{G}$ is a class of ${(n-o(n),g(n))}$-selective suppressors.\qed
\end{proof}

\medskip

The most natural question that arises by Definitions~\ref{amplifiers-def}
and~\ref{suppressors-def} is whether there exists any class of strong
selective amplifiers/suppressors, as well as for which functions $h(n)$ and $%
g(n)$ there exist classes of $g(n)$-universal amplifiers/suppressors and
classes of $(h(n),g(n))$-selective amplifiers/suppressors. In Section~\ref%
{amplifiers-sec} and~\ref{suppressors-sec} we provide our results on
amplifiers and suppressors, respectively.

\section{Amplifier bounds\label{amplifiers-sec}}

In this section we prove that there exist no strong universal amplifiers
(Section~\ref{no-universal-amplifiers-subsec}), although there exists a
class of strong selective amplifiers (Section~\ref%
{selective-amplifiers-subsec}).

\subsection{Non-existence of strong universal amplifiers\label%
{no-universal-amplifiers-subsec}}

\begin{theorem}
\label{no-universal-amplifiers-thm}
For any function $g(n)=\omega (n^{\frac{3}{4}})$
there exists no graph class $\mathcal{G}$ of $g(n)$-universal amplifiers for any $r>r_{0}=1$.
\end{theorem}

\begin{proof}
The proof is done by contradiction. 
Let $r_{0}=1$ and $g(n)=\omega (n^{\frac{3}{4}})$. 
Then $g(n)=\Omega (n^{\frac{3}{4}} \phi(n))$ for some function $\phi (n)=\omega (1)$, 
i.e.~$g(n)=\Omega (n^{1-\delta})$ where $\delta =\frac{1}{4} - \frac{\log \phi(n)}{\log n}$. 
Suppose that $\mathcal{G}$ is a class of $g(n)$%
-universal amplifiers. That is, for every graph ${G=(V,E)\in \mathcal{G}}$
with~${n\geq n_{0}}$ vertices, the fixation probability of $G$ is ${%
f_{r}(G)\geq 1-\frac{c(r)}{g(n)}\geq 1-\frac{c_{0}(r)}{n^{1-\delta }}}$ for
every ${r>1}$, where $c(r),c_{0}(r)$ are two functions that depend only on $%
r $. We partition the vertex set $V$ into three sets $V_{1},V_{2},V_{3}$
such that%
\begin{eqnarray}
V_{1} &=&\{v\in V:f_{r}(v)\geq {1-\frac{c_{0}(r)}{n^{1-\delta }}}\}
\label{V1-partition-eq} \\
V_{2} &=&\{v\in V\setminus V_{1}:f_{r}(v)\geq {1-\frac{c_{1}(r)}{%
n^{1-2\delta }}}\}  \label{V2-partition-eq} \\
V_{3} &=&\{v\in V\setminus V_{1}:f_{r}(v)<{1-\frac{c_{1}(r)}{n^{1-2\delta }}}%
\}  \label{V3-partition-eq}
\end{eqnarray}%
where $c_{1}(r)$ is an appropriate function of $r$ (to be specified below).
Note that $V_{1}\neq \emptyset $, since ${f_{r}(G)\geq 1-\frac{c_{0}(r)}{%
n^{1-\delta }}}$ by assumption. Using~(\ref{V3-partition-eq}), the fixation
probability $f_{r}(G)$ of~$G$~is upper-bounded by%
\begin{eqnarray}
f_{r}(G) &\leq &\frac{(|V_{1}|+|V_{2}|)\cdot 1+|V_{3}|\cdot (1-{\frac{%
c_{1}(r)}{n^{1-2\delta }}})}{n}  \notag \\
&=&1-\frac{|V_{3}|}{n}\cdot {\frac{c_{1}(r)}{n^{1-2\delta }}}
\label{universal-amplifier-eq}
\end{eqnarray}%
Now, since $f_{r}(G)\geq 1-{\frac{c_{0}(r)}{n^{1-\delta }}}$, it follows by (%
\ref{universal-amplifier-eq}) that $1-{\frac{c_{0}(r)}{n^{1-\delta }}}\leq 1-%
\frac{|V_{3}|}{n}\cdot {\frac{c_{1}(r)}{n^{1-2\delta }}}$, and thus%
\begin{equation}
|V_{3}|\leq n^{1-\delta }\frac{c_{0}(r)}{c_{1}(r)}  \label{V3-upper-bound-eq}
\end{equation}

For an arbitrary vertex $v\in V$, we obtain an upper bound on the
probability $f_{r}(v)$ by assuming that fixation is reached if the process
reaches at least two infected vertices, when it starts with only $v$ being
infected. Therefore%
\begin{equation}
f_{r}(v)\leq \frac{r\cdot 1+\sum_{x\in N(v)}\frac{1}{\deg x}\cdot 0}{%
r+\sum_{x\in N(v)}\frac{1}{\deg x}}=\frac{r}{r+\sum_{x\in N(v)}\frac{1}{\deg x}}  
\label{universal-f(v)-bound-eq}
\end{equation}%
for every $v\in V$. It follows now by~(\ref{V1-partition-eq}) and~(\ref%
{universal-f(v)-bound-eq}) that for every $v\in V_{1}$,%
\begin{eqnarray}
{1-\frac{c_{0}(r)}{n^{1-\delta }}} &\leq &\frac{r}{r+\sum_{x\in N(v)}\frac{1}{\deg x}}\Leftrightarrow   \notag \\
\sum_{x\in N(v)}\frac{1}{\deg x} &\leq &\frac{r\cdot c_{0}(r)}{n^{1-\delta
}-c_{0}(r)}\leq \frac{c^{\prime }(r)}{n^{1-\delta }}  \label{Sigma-for-V1-eq}
\end{eqnarray}%
for an appropriate function $c^{\prime }(r)$ of $r$. Therefore, since $%
\sum_{x\in N(v)}\frac{1}{\deg x}\geq \deg (v)\cdot \frac{1}{n}$,~(\ref%
{Sigma-for-V1-eq})~implies that%
\begin{equation}
\deg v\leq c^{\prime }(r)\cdot n^{\delta }  \label{upper-bound-deg-V1-eq}
\end{equation}%
for every $v\in V_{1}$. Furthermore, since $\frac{1}{\deg u}\leq \sum_{x\in
N(v)}\frac{1}{\deg x}$ for every $u\in N(v)$,~(\ref{Sigma-for-V1-eq})
implies that%
\begin{equation}
\deg u\geq n^{1-\delta} \frac{1}{c^{\prime }(r)}
\label{lower-bound-deg-N(V1)-eq}
\end{equation}%
for every $u\in N(v)$, where $v\in V_{1}$. Define now the value of the
function $c_{1}(r)$ (cf. (\ref{V2-partition-eq})-(\ref{V3-partition-eq})) as:%
\begin{equation*}
c_{1}(r)=2c_{0}(r)\cdot c^{\prime }(r)
\end{equation*}%
and thus (\ref{V3-upper-bound-eq}) becomes%
\begin{equation}
|V_{3}|\leq n^{1-\delta }\frac{1}{2c^{\prime }(r)}
\label{V3-upper-bound-eq-2}
\end{equation}%
Similarly to (\ref{Sigma-for-V1-eq}), it follows by~(\ref{V2-partition-eq})
and~(\ref{universal-f(v)-bound-eq}) that for every $v\in V_{2}$,%
\begin{equation*}
\sum_{u\in N(v)}\frac{1}{\deg u}\leq \frac{r\cdot c_{1}(r)}{n^{1-2\delta
}-c_{1}(r)}\leq \frac{c^{\prime \prime }(r)}{n^{1-2\delta }}
\end{equation*}%
for some function $c^{\prime \prime }(r)$ of $r$, and that%
\begin{equation}
\deg v\leq c^{\prime \prime }(r)\cdot n^{2\delta }
\label{upper-bound-deg-V2-eq}
\end{equation}%
for every $v\in V_{2}$.

In the remainder of the proof we will use a generic upper bound on the 
fixation probability of undirected graphs which was proved in~\cite{MertziosNRS-TCS2013}. 
To state this generic upper bound, we need to define for every vertex $v$ 
the quantity $Q_{v}=\sum_{x\in N(v)}\frac{1}{\deg (x)}$, 
as well as for every edge $uv \in E$ the quantity 
$Q_{uv}=\sum_{x\in N(v)\setminus \{u\}}\frac{1}{\deg (x)}%
+\sum_{x\in N(u)\setminus \{v\}}\frac{1}{\deg (x)}$.
Intuitively, $Q_{v}$ is an indicator of how ``strong'' the neighbors of $v$ are, compared to $v$. 
Roughly, if $Q_{v}$ is larger, then $v$ is weak compared to its neighbors. 
Similarly, $Q_{uv}$ is an indicator of how ``strong'' a pair of neighbored vertices $u,v$ are, 
compared to their neighborhood $N(u) \cup N(v) \setminus \{u,v\}$.

Let $v\in V$ such that $f_{r}(v)$ is maximized. 
The generic upper bound of~\cite{MertziosNRS-TCS2013} on the fixation probability is\footnote{In the original statement of this upper bound (see Theorem~1 in~\cite{MertziosNRS-TCS2013}), 
the maximization is taken over all edges $uv\in E$. However, it follows from the proof of Theorem~1 in~\cite{MertziosNRS-TCS2013} that the maximization can actually be taken over all edges $uv$ such that $f_{r}(v)$ is maximized, i.e.~as stated in the inequality (\ref{f-G-upper-bound-eq-0}).}
\begin{equation}
f_{r}(G)\leq \max_{u\in N(v)}\frac{r^{2}}{r^{2}+rQ(v)+\frac{Q_{v}Q_{uv}}{2}}.
\label{f-G-upper-bound-eq-0}
\end{equation}%
The main idea for the proof of (\ref{f-G-upper-bound-eq-0}) is to construct an auxiliary Markov chain $\widetilde{\mathcal{M}}$, in which the probability of reaching a specific absorbing state is at least 
as large as the fixation probability in the original Markov chain. To favor fixation in this auxiliary Markov chain $\widetilde{\mathcal{M}}$, fixation is assumed to be reached whenever we reach three mutants in the population. For further details on this upper bound on $f_{r}(G)$ we refer to Theorem~1 in~\cite{MertziosNRS-TCS2013}.

In the following let $v\in V$ such that $f_{r}(v)$ is maximized. Then clearly $v\in V_{1}$. Furthermore, (\ref{f-G-upper-bound-eq-0}) implies that%
\begin{equation}
f_{r}(G)\leq \max_{u\in N(v)}\frac{2r^{2}}{2r^{2}+Q_{v}Q_{uv}}.
\label{f-G-upper-bound-eq-1}
\end{equation}%
Let $u_{0}\in N(v)$ be such that the right hand side of~(\ref{f-G-upper-bound-eq-1})~is maximized, and thus 
\begin{equation}
f_{r}(G)\leq \frac{2r^{2}}{2r^{2}+Q_{v}Q_{u_{0}v}}
\label{f-G-upper-bound-eq-2}
\end{equation}%
To arrive to a contradiction to our assumption on $\delta$, first we upper-bound the product $Q_{v}Q_{u_{0}v}$ by a quantity proportional to $\frac{1}{n^{1-\delta}}$, and then we lower-bound 
$Q_{v}Q_{u_{0}v}$ by a quantity proportional to~$n^{-3\delta}$ (see inequality (\ref{upper-lower-bound-QuQuv-eq}) below). 
Since ${f_{r}(G)\geq 1-\frac{c_{0}(r)}{n^{1-\delta }}}$, it
follows by~(\ref{f-G-upper-bound-eq-2}) that%
\begin{eqnarray}
{1-\frac{c_{0}(r)}{n^{1-\delta }}} &\leq &\frac{2r^{2}}{2r^{2}+Q_{v}Q_{uv}}\Leftrightarrow  \notag \\
n^{1-\delta }Q_{v}Q_{u_{0}v} &\leq
&2r^{2}c_{0}(r)+c_{0}(r)Q_{v}Q_{u_{0}v}\Leftrightarrow  \notag \\
Q_{v}Q_{u_{0}v} &\leq &\frac{2r^{2}c_{0}(r)}{n^{1-\delta }-c_{0}(r)}\leq 
\frac{c^{\prime \prime \prime }(r)}{n^{1-\delta }}
\label{upper-bound-QuQuv-eq}
\end{eqnarray}%
for an appropriate function $c^{\prime \prime \prime }(r)$ of $r$.

Since $v \in V_{1}$ and $u_{0}\in N(v)$, it follows by (\ref{lower-bound-deg-N(V1)-eq}) and~(\ref{V3-upper-bound-eq-2}) 
that $u_{0}$ has at least $n^{1-\delta }\frac{1}{2c^{\prime }(r)}$
neighbors in $V_{1}\cup V_{2}$. Thus (\ref{upper-bound-deg-V1-eq}) and~(\ref%
{upper-bound-deg-V2-eq}) imply that 
\begin{equation*}
\sum_{x\in N(u_{0})\setminus \{v\}}\frac{1}{\deg (x)}\geq \left(
n^{1-\delta }\frac{1}{2c^{\prime }(r)}-1\right) \cdot \min \left\{ \frac{1}{%
c^{\prime }(r)\cdot n^{\delta }},\frac{1}{c^{\prime \prime }(r)\cdot
n^{2\delta }}\right\} =\Omega (n^{1-3\delta })
\end{equation*}%
Furthermore $Q_{u_{0}v}\geq \sum_{x\in N(u_{0})\setminus \{v\}}\frac{1}{\deg
(x)}$ by the definition of $Q_{u_{0}v}$, and thus $Q_{u_{0}v}=\Omega
(n^{1-3\delta })$. Moreover $Q_{v}Q_{u_{0}v}=\Omega (n^{-3\delta })$, since $%
Q_{v}=\Omega (\frac{1}{n})$. Therefore it follows by~(\ref%
{upper-bound-QuQuv-eq}) that%
\begin{equation}
\Omega (n^{-3\delta })=Q_{v}Q_{u_{0}v}\leq \frac{c^{\prime \prime \prime }(r)%
}{n^{1-\delta }}.
\label{upper-lower-bound-QuQuv-eq}
\end{equation}%
This is a contradiction, since $\delta =\frac{1}{4} - \frac{\log \phi(n)}{\log n}$ by assumption, 
where $\phi(n)=\omega(1)$.
Therefore there exists no class $\mathcal{G}$ of $g(n)$-universal amplifiers
for any $r>r_{0}=1$, where $g(n)=\omega (n^\frac{3}{4})$.\qed
\end{proof}

\medskip

The next corollary follows from Theorem~\ref{no-universal-amplifiers-thm}.

\begin{corollary}
\label{no-strong-universal-amplifiers-cor}There exists no infinite class $%
\mathcal{G}$ of undirected graphs which are strong universal amplifiers.
\end{corollary}

\subsection{A class of strong selective amplifiers\label%
{selective-amplifiers-subsec}}

In this section we present the first class $\mathcal{G}=\{G_{n}:n\geq 1\}$
of strong selective amplifiers, which we call the \emph{urchin} graphs.
Namely, the graph $G_{n}$ has $2n$ vertices, consisting of a clique with $n$
vertices, an independent set of $n$ vertices, and a perfect matching between
the clique and the independent set, as illustrated in Figure~\ref%
{urchin-fig}. For every graph $G_{n}$, we refer for simplicity to a vertex
of the clique of $G_{n}$ as a \emph{clique vertex} of $G_{n}$, and to a
vertex of the independent set of $G_{n}$ as a \emph{nose} of $G_{n}$,
respectively. We prove in this section that the class $\mathcal{G}$ of
urchin graphs are strong selective amplifiers. Namely, we prove that,
whenever $r>r_{0}=5$, the fixation probability of any nose $v$ of any graph $%
G_{n}$ is $f_{r}(v)\geq 1-\frac{c(r)}{n}$, where $c(r)$ is a function that
depends only on the fitness $r$ of the mutant.

\begin{figure}[htb]
\centering\includegraphics[scale=0.68]{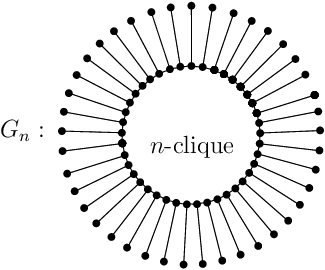}
\caption{The ``urchin'' graph $G_{n}$ with $2n$ vertices.}
\label{urchin-fig}
\end{figure}

\begin{figure}[tbh]
\centering%
\subfigure[]{ \label{q-i-x-k-state-fig}
\includegraphics[scale=0.68]{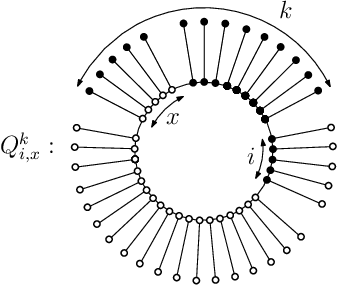}} \hspace{0.1cm} 
\subfigure[]{ \label{p-i-k-state-fig-1}
\includegraphics[scale=0.68]{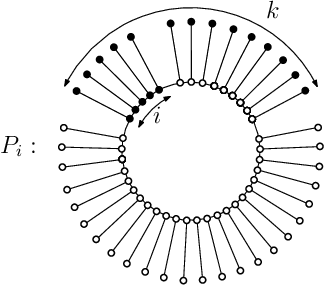}} \hspace{0.1cm} 
\subfigure[]{ \label{p-i-k-state-fig-2}
\includegraphics[scale=0.68]{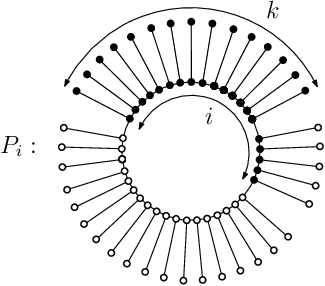}}
\caption{The state (a)~$Q_{i,x}^{k}$ and the state $P_{i}^{k}$, where (b)~$%
i\leq k$, and~(c)~$i>k$.}
\label{states-k-fig}
\end{figure}

Let $v$ be a clique vertex (resp.~a nose) and $u$ be its adjacent nose
(resp.~clique vertex). Let $v$ be infected; if $u$ is not infected, then $v$
is called an \emph{isolated clique vertex} (resp.~\emph{isolated nose}),
otherwise $v$ is called a \emph{covered clique vertex} (resp.~\emph{covered
nose}). Let $k\in \{0,1,\ldots ,n\}$, $i\in \{0,1,2,\ldots ,n-k\}$, and $%
x\in \{0,1,2,\ldots ,k\}$. Denote by $Q_{i,x}^{k}$ the state of $G_{n}$ with
exactly $i$ isolated clique vertices, $x$ isolated noses, and $k-x$ covered
noses. An example of the state $Q_{i,x}^{k}$ is illustrated in Figure~\ref%
{q-i-x-k-state-fig}. Furthermore, for every $k,i\in \{0,1,\ldots ,n\}$, we
define the state $P_{i}^{k}$ of $G_{n}$ as follows. If $i\leq k$, then $%
P_{i}^{k}$ is the state with exactly $i$ covered noses and $k-i$ isolated
noses. If $i>k$, then $P_{i}^{k}$ is the state with exactly $k$ covered
noses and $i-k$ isolated clique vertices. Note that $Q_{i,0}^{k}=P_{k+i}^{k}$
and $Q_{0,x}^{k}=P_{k-x}^{k}$, for every $k\in \{0,1,\ldots ,n\}$, $i\in
\{0,1,2,\ldots ,n-k\}$, and $x\in \{0,1,2,\ldots ,k\}$. Two examples of the
state $P_{i}^{k}$, for the cases where $i\leq k$ and $i>k$, are illustrated
in Figures~\ref{p-i-k-state-fig-1} and~\ref{p-i-k-state-fig-2}, respectively.

Let $k\in \{1,2,\ldots ,n-1\}$. For all appropriate values of $i$ and $x$,
we denote by $q_{i,x}^{k}$ (resp.~$p_{i}^{k}$) the probability that,
starting at state $Q_{i,x}^{k}$ (resp.~$P_{i}^{k}$) we eventually arrive to
a state with $k+1$ infected noses before we arrive to a state with $k-1$
infected noses.

\begin{lemma}
\label{domination-1-lem}Let $k\in \{1,2,\ldots ,n-1\}$. Then $%
q_{i,x}^{k}>q_{i-1,x-1}^{k}$, for every $i\in \{1,2,\ldots ,n-k\}$ and every 
$x\in \{1,2,\ldots ,k\}$.
\end{lemma}

\begin{proof}
Denote by $\mathcal{M}_{1}$ the Markov chain with starting state $%
Q_{i,x}^{k} $. Similarly, denote by $\mathcal{M}_{2}$ the Markov chain with
starting state $Q_{i-1,x-1}^{k}$. Note that, initially, both Markov chains $%
\mathcal{M}_{1}$ and $\mathcal{M}_{2}$ have the same number $2(k-x)+x+i$ of
infected vertices. Moreover, $\mathcal{M}_{1}$ and $\mathcal{M}_{2}$
coincide initially on all their vertices except two. In particular, $%
\mathcal{M}_{1}$ has initially an isolated nose $u$, which is a covered nose
in $\mathcal{M}_{2}$. Furthermore, $\mathcal{M}_{1}$ has initially an
isolated clique vertex $v$, which is an uninfected clique vertex in $%
\mathcal{M}_{2}$. Denote by $u^{\prime }$ the (unique) clique vertex that is
adjacent to $u$ in $G_{n}$. Furthermore denote by $v^{\prime }$ the (unique)
nose that is adjacent to $v$ in $G_{n}$. Note that, initially, $u^{\prime }$
is uninfected in $\mathcal{M}_{1}$ and infected in $\mathcal{M}_{2}$, while $%
v^{\prime }$ is uninfected in both $\mathcal{M}_{1}$ and $\mathcal{M}_{2}$.

Note that at every iteration of the processes $\mathcal{M}_{1}$ and $%
\mathcal{M}_{2}$, one vertex $w$ is activated and then it replaces a
neighbor $w^{\prime }$ of it by an offspring of $w$. Thus, an equivalent way
to analyze these processes is to consider that, at every iteration, one
directed edge between two adjacent vertices is activated (with the
appropriate probability). In order to prove that $%
q_{i,x}^{k}>q_{i-1,x-1}^{k} $, we simulate the progress of $\mathcal{M}_{1}$
by the random choices made at the corresponding steps by $\mathcal{M}_{2}$.
In particular, we simulate the processes $\mathcal{M}_{1}$ and $\mathcal{M}%
_{2}$ until they reach states $S_{1}$ and $S_{2}$, respectively, such that
either $S_{1}=S_{2}$, or one of $S_{1}$ and $S_{2}$ is strictly included in
the other. Furthermore, during the whole simulation of $\mathcal{M}_{1}$ by $%
\mathcal{M}_{2}$, before we reach such states $S_{1}$ and $S_{2}$, both $%
\mathcal{M}_{1}$ and $\mathcal{M}_{2}$ have the same number of infected
vertices at the corresponding iterations.

Suppose that a vertex $w$ is activated for reproduction in $\mathcal{M}_{2}$ 
and that it places its offspring at a vertex $w'$. 
If both $w,w' \notin \{u,u^{\prime },v,v^{\prime }\}$ then we mimic this step in $\mathcal{M}_{1}$, 
i.e.~in $\mathcal{M}_{1}$ we also activate $w$ for reproduction and we place its offspring at $w'$. 
In the next three paragraphs we consider each of the remaining cases where at least one of 
$w,w'$ belongs to $\{u,u^{\prime },v,v^{\prime }\}$. For each of these cases we specify 
how we simulate this step in $\mathcal{M}_{1}$.

If the clique vertex $v$ (resp.~$u^{\prime }$) is activated for reproduction
in $\mathcal{M}_{2}$, then we activate $u^{\prime }$ (resp.~$v$) in $%
\mathcal{M}_{1}$. In this case, if $v$ (resp.~$u^{\prime }$) places in $%
\mathcal{M}_{2}$ its offspring at a clique vertex $w\neq u^{\prime }$ (resp.~%
$w\neq v$), then $u^{\prime }$ (resp.~$v$) places in $\mathcal{M}_{1}$ its
offspring at the same clique vertex $w$. If $v$ (resp.~$u^{\prime }$) places
in $\mathcal{M}_{2}$ its offspring at the clique vertex $u^{\prime }$ (resp.~%
$v$), then $u^{\prime }$ (resp.~$v$) places in $\mathcal{M}_{1}$ its
offspring at the clique vertex $v$ (resp.~$u^{\prime }$); in this case we
arrive to two identical states in both $\mathcal{M}_{1}$ and $\mathcal{M}%
_{2} $. Finally, if $v$ (resp.~$u^{\prime }$) places in $\mathcal{M}_{2}$
its offspring at its adjacent nose $v^{\prime }$ (resp.~$u$), then $%
u^{\prime }$ (resp.~$v$) places in $\mathcal{M}_{1}$ its offspring at its
adjacent nose $u $ (resp.~$v^{\prime }$); in this case we arrive in $%
\mathcal{M}_{1}$ to a state, in which the infected vertices are a strict
subset (resp.~superset) of the infected vertices in $\mathcal{M}_{2}$.

If a clique vertex $w\notin \{v,u^{\prime }\}$ is activated for reproduction
in $\mathcal{M}_{2}$, and if $w$ places in $\mathcal{M}_{2}$ its offspring
at $v$ (resp.~$u^{\prime }$), then $w$ places in $\mathcal{M}_{1}$ its
offspring at $u^{\prime }$ (resp.~$v$). In this case, if the number of
infected vertices changes in $\mathcal{M}_{2}$, then we arrive to the same
state in both $\mathcal{M}_{1}$ and $\mathcal{M}_{2}$.

Finally, if the nose $w=v^{\prime }$ (resp.~$w=u$) is activated for
reproduction in $\mathcal{M}_{2}$, then we activate the same nose also in $%
\mathcal{M}_{1}$. In this case we arrive in $\mathcal{M}_{1}$ to a state, in
which the infected vertices are a strict subset (resp.~superset) of the
infected vertices in $\mathcal{M}_{2}$.

Note now that $q_{i,x}^{k}>q_{i-1,x-1}^{k}$ if and only if, in the above
simulation of $\mathcal{M}_{1}$ by $\mathcal{M}_{2}$, the probability that
we arrive to strictly more infected vertices in $\mathcal{M}_{1}$ than $%
\mathcal{M}_{2}$ is greater or equal to the probability that we arrive to
strictly less infected vertices in $\mathcal{M}_{1}$ than $\mathcal{M}_{2}$.
Furthermore, note that we arrive in $\mathcal{M}_{1}$ to a state with
strictly more or strictly less infected vertices than in $\mathcal{M}_{2}$
only when one of the edges $uu^{\prime }$ or $vv^{\prime }$ is activated (in
some direction) in the process $\mathcal{M}_{2}$. In particular, whenever
this event occurs, $\mathcal{M}_{1}$ receives strictly more infected
vertices than $\mathcal{M}_{2}$, if either $u$ places its offspring at $%
u^{\prime }$ in $\mathcal{M}_{2}$, or if $u^{\prime }$ places its offspring
at $u$ in $\mathcal{M}_{2}$. Similarly, $\mathcal{M}_{1}$ receives strictly
less infected vertices than $\mathcal{M}_{2}$, if either $v$ places its
offspring at $v^{\prime }$ in $\mathcal{M}_{2}$, or if $v^{\prime }$ places
its offspring at $v$ in $\mathcal{M}_{2}$. The ratio of these probabilities
is%
\begin{equation*}
\frac{r\cdot \frac{1}{n}+r\cdot 1}{\frac{1}{n}+1}=r>1
\end{equation*}%
and thus $q_{i,x}^{k}>q_{i-1,x-1}^{k}$.\qed
\end{proof}

\begin{corollary}
\label{domination-2-cor}Let $k\in \{1,2,\ldots ,n-1\}$, $i\in \{0,1,\ldots
,n-k\}$, and $x\in \{0,1,\ldots ,k\}$. Then $q_{i,x}^{k}>p_{k+i-x}^{k}$.
\end{corollary}

\begin{proof}
Suppose first that $i\geq x$. Then Lemma~\ref{domination-1-lem} implies that 
$q_{i,x}^{k}>q_{i-1,x-1}^{k}>\ldots >q_{i-x,0}^{k}=p_{k+i-x}^{k}$. Suppose
now that $i<x$. Then Lemma~\ref{domination-1-lem} implies that $%
q_{i,x}^{k}>q_{i-1,x-1}^{k}>\ldots >q_{0,x-i}^{k}=p_{k+i-x}^{k}$.\qed
\end{proof}

\medskip

Now, starting from the Markov chain of the generalized Moran process, we define the Markov chain $\mathcal{M}$ by replacing any transition to a state $Q_{i,x}^{k}$ with a transition to state $P_{k+i-x}^{k}$. 
Then, for every nose~$v$ of the graph $G_{n}$, the fixation probability $f_{r}(v)$ of $v$ is by Corollary~\ref{domination-2-cor} greater than or equal to the fixation probability of state $P^{1}_{0}$ 
in the Markov chain $\mathcal{M}$. 
Thus, in order to compute a lower bound on the fixation probability $f_{r}(v)$ of a nose $v$ in $G_{n}$, we will compute a lower bound on the fixation probability of state $P^{1}_{0}$ in $\mathcal{M}$ 
(cf.~Theorem~\ref{birth-death-lower-bound-thm}).

\begin{figure}[tbh]
\centering\includegraphics[scale=0.68]{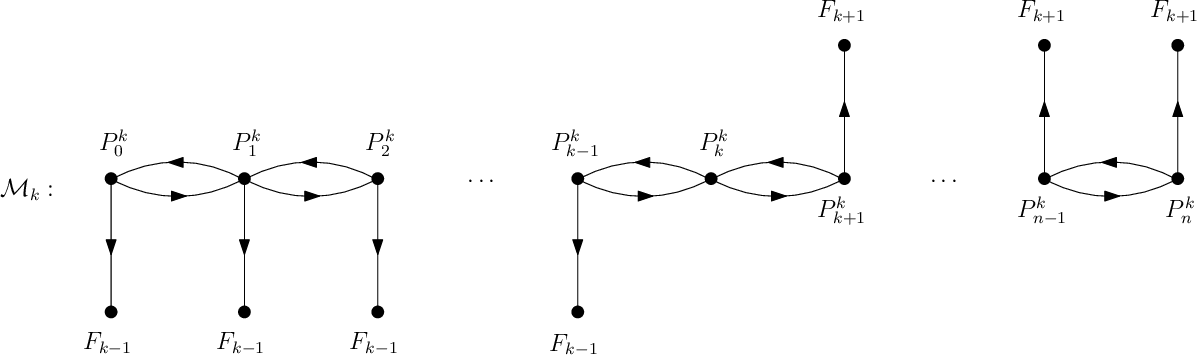}
\caption{The state graph of the relaxed Markov chain $\mathcal{M}_{k}$,
where $k\in \{1,2,\ldots ,n-1\}$.}
\label{state-graph-M-k-fig}
\end{figure}

In order to analyze $\mathcal{M}$, we decompose it first into the $n-1$
smaller Markov chains $\mathcal{M}_{1},\mathcal{M}_{2},\ldots ,\mathcal{M}%
_{n-1}$, as follows. For every $k\in \{1,2,\ldots ,n-1\}$, the Markov chain $%
\mathcal{M}_{k}$ captures all transitions of $\mathcal{M}$ between states
with $k$ infected noses. The state graph of $\mathcal{M}_{k}$ is illustrated
in Figure~\ref{state-graph-M-k-fig}, where we denote by $F_{k-1}$ (resp.~$%
F_{k+1}$) an \emph{arbitrary} state with $k-1$ (resp.~$k+1$) infected noses.
Moreover, we consider $F_{k-1}$ and $F_{k+1}$ as absorbing states of $%
\mathcal{M}_{k}$. Since we want to compute a lower bound of the fixation
probability, whenever we arrive at state $F_{k+1}$ (resp.~at state $F_{k-1}$%
), we assume that we have the smallest number of infected clique vertices
with $k+1$ (resp.~with $k-1$) infected noses. That is, whenever $\mathcal{M}%
_{k}$ reaches state $F_{k+1}$, we assume that $\mathcal{M}$ has reached
state $P_{k+1}^{k+1}$ (and thus we move to the Markov chain $\mathcal{M}%
_{k+1}$). Similarly, whenever $\mathcal{M}_{k}$ reaches state $F_{k-1}$, we
assume that $\mathcal{M}$ has reached state $P_{0}^{k-1}$ (and thus we move
to the Markov chain~$\mathcal{M}_{k-1}$).

\subsubsection{A decomposition of $\mathcal{M}_{k}$ into two Markov chains 
\label{two-Markov-chains-subsubsec}}

In order to analyze the Markov chain $\mathcal{M}_{k}$, where $k\in
\{1,2,\ldots ,n-1\}$, we decompose it into two smaller Markov chains $\{%
\mathcal{M}_{k}^{1},\mathcal{M}_{k}^{2}\}$, as they are shown in Figure~\ref%
{state-graph-M-k-1-2-fig}.

\begin{figure}[tbh]
\centering%
\subfigure[]{ \label{state-graph-M-k-1-fig}
\includegraphics[scale=0.68]{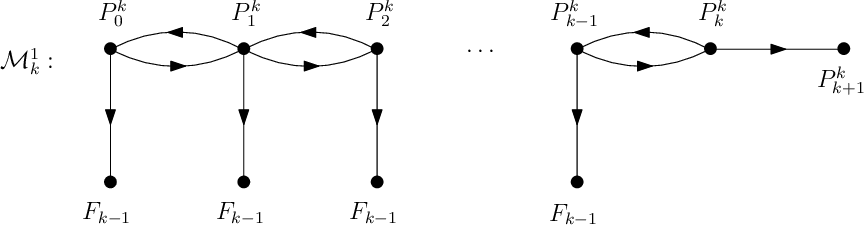}} \hspace{0.1cm} 
\subfigure[]{ \label{state-graph-M-k-2-fig}
\includegraphics[scale=0.68]{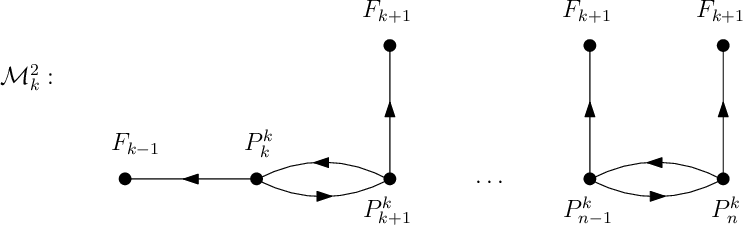}}
\caption{The two Markov chains $\mathcal{M}_{k}^{1}$ and $\mathcal{M}%
_{k}^{2} $, where $k\in \{1,2,\ldots ,n-1\}$.}
\label{state-graph-M-k-1-2-fig}
\end{figure}

In $\mathcal{M}_{k}^{1}$, we consider the state $P_{k+1}^{k}$ absorbing. For
every $i\in \{0,1,\ldots ,k\}$ denote by $h_{i}^{k}$ the probability that,
starting at state $P_{i}^{k}$ in $\mathcal{M}_{k}^{1}$, we eventually reach
state $P_{k+1}^{k}$ before we reach state $F_{k-1}$, cf.~Figure~\ref%
{state-graph-M-k-1-fig}. In this Markov chain $\mathcal{M}_{k}^{1}$, every
transition probability between two states is equal to the corresponding
transition probabilities in $\mathcal{M}_{k}$.

In $\mathcal{M}_{k}^{2}$, we denote by $s_{i}^{k}$, where $i\in
\{k,k+1,\ldots ,n\}$, the probability that starting at state $P_{i}^{k}$ we
eventually reach state $F_{k+1}$ before we reach state $F_{k-1}$, cf.~Figure~%
\ref{state-graph-M-k-2-fig}. In this Markov chain $\mathcal{M}_{k}^{2}$, the
transition probability from state $P_{k}^{k}$ to state $P_{k+1}^{k}$
(resp.~to state $F_{k-1}$) is equal to $h_{k}^{k}$ (resp.~$1-h_{k}^{k}$),
while all other transition probabilities between two states in $\mathcal{M}%
_{k}^{2}$ are the same as the corresponding transition probabilities in $%
\mathcal{M}_{k}$.

In order to prove the main result of this section, 
namely that the class $\mathcal{G}=\{G_{n}:n\geq 1\}$ of urchin graphs is a class of strong 
selective amplifiers (cf.~Theorem~\ref{urchin-strong-selective-amplifiers-thm}), we first need to prove a series of technical results which can be outlined as follows. 

First, we compute (cf.~Lemmas~\ref{hk-statement-lem} and~\ref{h-0-k-statement-lem}) 
a lower bound on the probability $h_{0}^{k}$. That is, we compute a lower bound on the probability that, 
starting at state $P_{0}^{k}$ in $\mathcal{M}_{k}^{1}$ 
(i.e.~starting with exactly $k$ infected noses and all other vertices uninfected), 
we eventually infect at least $k+1$ clique vertices before disinfecting any infected nose.
Second, we compute by Lemma~\ref{sk-statement-lem} a lower bound on the probability $s_{k}^{k}$. 
That is, we compute a lower bound on the probability that, 
starting at state~$P_{k}^{k}$ in $\mathcal{M}_{k}^{2}$ 
(i.e.~starting with $k$ noses and their $k$ adjacent clique vertices infected, 
while all other vertices are uninfected), we eventually infect one more nose before disinfecting any infected nose.

Then, using these lower bounds for $h_{0}^{k}$ and $s_{k}^{k}$, we are able to compute a lower bound on 
the  fixation probability of state $P_{0}^{1}$ in the Markov chain $\mathcal{M}$ 
(cf.~Figure~\ref{state-graph-M-full-fig}). 
To do so, we further relax the Markov chain $\mathcal{M}$ to another Markov chain $\mathcal{M}'$, 
which is equivalent to a birth-death process~$\mathcal{B}_{n}$ 
(cf.~Figures~\ref{state-graph-M-prime-full-fig-1} and~\ref{state-graph-M-prime-full-fig-2}, respectively). 
As it turns out, the fixation probability $p_{1}$ of state~$P_{0}^{1}$ in this birth-death 
process $\mathcal{B}_{n}$ (cf.~Figure~\ref{state-graph-M-prime-full-fig-2}) 
is a lower bound on the fixation probability of state~$P_{0}^{1}$ in the Markov chain $\mathcal{M}$ (cf.~Figure~\ref{state-graph-M-full-fig}), which is in turn a lower bound on the 
fixation probability~$f_{r}(v)$ of a nose $v$ in the urchin graph $G_{n}$. 
Finally we prove in Theorem~\ref{birth-death-lower-bound-thm} that $p_{1}\geq 1-\frac{c(r)}{n}$, 
for some appropriate function $c(r)$ of $r$, and thus the same lower bound also applies to 
the fixation probability~$f_{r}(v)$ of a nose $v$ in $G_{n}$ (cf.~Theorem~\ref{urchin-strong-selective-amplifiers-thm}).

\begin{lemma}
\label{hk-statement-lem}In the Markov chain $\mathcal{M}_{k}^{1}$, for any $%
r>1$,%
\begin{equation}
h_{k}^{k}\geq 1-\frac{2}{n(r-1)+1}=1-O(\frac{1}{n})  \label{hk-statement}
\end{equation}
\end{lemma}

\begin{proof}
For $i=0$, the value of $h_{i}^{k}$ in the Markov chain $\mathcal{M}_{k}^{1}$
is%
\begin{equation}
h_{0}^{k}=\frac{rk\cdot h_{1}^{k}+\frac{k}{n}\cdot 0}{rk+\frac{k}{n}}=\frac{%
rn}{rn+1}\cdot h_{1}^{k}  \label{h0-eq-1}
\end{equation}%
and thus%
\begin{equation}
h_{1}^{k}-h_{0}^{k}=\frac{1}{rn}h_{0}^{k}\leq \frac{1}{rn}  \label{h0-eq-2}
\end{equation}%
Furthermore, for every $i\in \{1,\ldots ,k\}$, where $1\leq k\leq n-1$, the
value of $h_{i}^{k}$ in $\mathcal{M}_{k}^{1}$ can be computed as follows.%
\begin{equation}
h_{i}^{k}=\alpha _{i}^{k}\cdot h_{i+1}^{k}+\beta _{i}^{k}\cdot
h_{i-1}^{k}+\gamma _{i}^{k}\cdot 0  \label{hi-eq-1}
\end{equation}%
where%
\begin{eqnarray}
\alpha _{i}^{k} &=&\frac{r\frac{((k-i)n+i(n-i))}{n}}{\Sigma _{i}^{k}}  \notag
\\
\beta _{i}^{k} &=&\frac{\frac{i(n-i)}{n}}{\Sigma _{i}^{k}}
\label{alpha-beta-gamma-hi-eq} \\
\gamma _{i}^{k} &=&\frac{\frac{k-i}{n}}{\Sigma _{i}^{k}}  \notag
\end{eqnarray}%
and $\Sigma _{i}^{k}=r\frac{((k-i)n+i(n-i))}{n}+\frac{i(n-i)}{n}+\frac{k-i}{n%
}$. Therefore~(\ref{hi-eq-1}) implies that%
\begin{equation}
h_{i+1}^{k}-h_{i}^{k}=\frac{\beta _{i}^{k}}{\alpha _{i}^{k}}%
(h_{i}^{k}-h_{i-1}^{k})+\frac{\gamma _{i}^{k}}{\alpha _{i}^{k}}h_{i}^{k}
\label{hi-eq-2}
\end{equation}%
Furthermore,~(\ref{alpha-beta-gamma-hi-eq}) implies that%
\begin{eqnarray}
\frac{\beta _{i}^{k}}{\alpha _{i}^{k}} &=&\frac{1}{r}\cdot \frac{i(n-i)}{%
(k-i)n+i(n-i)}\leq \frac{1}{r}  \label{beta-over-alpha-hi-eq} \\
\frac{\gamma _{i}^{k}}{\alpha _{i}^{k}} &=&\frac{1}{r}\cdot \frac{k-i}{%
(k-i)n+i(n-i)}=\frac{1}{r}\cdot \frac{1}{n+\frac{i(n-i)}{k-i}}\leq \frac{1}{%
rn}  \label{gamma-over-alpha-hi-eq}
\end{eqnarray}%
Note that the inequality $\frac{\gamma _{i}^{k}}{\alpha _{i}^{k}}\leq \frac{1%
}{rn}$ in~(\ref{gamma-over-alpha-hi-eq}) holds also for $i=k$, since $\gamma
_{k}^{k}=0$. Therefore it follows by~(\ref{hi-eq-2}),~(\ref%
{beta-over-alpha-hi-eq}), and~(\ref{gamma-over-alpha-hi-eq}) that 
\begin{equation}
h_{i+1}^{k}-h_{i}^{k}\leq \frac{1}{r}(h_{i}^{k}-h_{i-1}^{k})+\frac{1}{rn}%
h_{i}^{k}  \label{hi-eq-3}
\end{equation}%
Thus, since $h_{k+1}^{k}=1$ by definition, it follows by~(\ref{hi-eq-3}) for 
$i=k$ that 
\begin{eqnarray*}
1-h_{k}^{k} &\leq &\frac{1}{r}(h_{k}^{k}-h_{k-1}^{k})+\frac{1}{rn}h_{k}^{k}
\\
&\leq &\ldots \\
&\leq &\frac{1}{r^{k}}(h_{1}^{k}-h_{0}^{k})+\frac{1}{rn}(h_{k}^{k}+\frac{1}{r%
}h_{k-1}^{k}+\ldots +\frac{1}{r^{k-1}}h_{1}^{k})
\end{eqnarray*}%
Therefore, since $h_{1}^{k}\leq h_{2}^{k}\leq \ldots \leq h_{k}^{k}$ and $%
h_{1}^{k}-h_{0}^{k}\leq \frac{1}{rn}$ by~(\ref{h0-eq-2}), it follows that%
\begin{equation*}
1-h_{k}^{k} \leq \frac{1}{r^{k+1}n}+\frac{1}{n(r-1)}h_{k}^{k}\leq \frac{1}{(r-1)n}+\frac{1%
}{n(r-1)}h_{k}^{k}
\end{equation*}%
Therefore%
\begin{equation}
h_{k}^{k}\geq 1-\frac{2}{n(r-1)+1}  \label{hk-eq}
\end{equation}\qed
\end{proof}

\begin{lemma}
\label{h-0-k-statement-lem}In the Markov chain $\mathcal{M}_{k}^{1}$, for
any $r>1$,%
\begin{equation}
h_{0}^{k}\geq 1-\frac{k+2}{n(r-1)}  \label{hi-eq-4}
\end{equation}
\end{lemma}

\begin{proof}
Let $1\leq i\leq k$. Recall by~(\ref{hi-eq-3}) in the proof of Lemma~\ref%
{hk-statement-lem} that%
\begin{equation*}
h_{i+1}^{k}-h_{i}^{k} \leq \frac{1}{r}(h_{i}^{k}-h_{i-1}^{k})+\frac{1}{rn}%
h_{i}^{k} \leq \ldots \leq \frac{1}{r^{i}}(h_{1}^{k}-h_{0}^{k})+\frac{1}{n(r-1)}
\end{equation*}%
Therefore it follows by~(\ref{h0-eq-2}) that%
\begin{equation}
h_{i+1}^{k}-h_{i}^{k}\leq \frac{1}{r^{i+1}n}+\frac{1}{n(r-1)}
\label{hi-eq-5}
\end{equation}%
Summing up~(\ref{hi-eq-5}) for every $i=1,2,\ldots ,k$, it follows that%
\begin{eqnarray*}
1-h_{1}^{k} &\leq &\frac{1}{n}(\frac{1}{r^{2}}+\ldots +\frac{1}{r^{k+1}})+%
\frac{k}{n(r-1)} \\
&\leq &\frac{1}{n}\frac{1}{r(r-1)}+\frac{k}{n(r-1)}\leq \frac{k+1}{n(r-1)}
\end{eqnarray*}%
since $h_{k+1}^{k}=1$, and thus%
\begin{equation}
h_{1}^{k}\geq 1-\frac{k+1}{n(r-1)}  \label{hi-eq-6}
\end{equation}%
Therefore it follows now by~(\ref{h0-eq-1}) that%
\begin{equation*}
h_{0}^{k} \geq \frac{rn}{rn+1}(1-\frac{k+1}{n(r-1)})\geq \frac{n(r-1)}{%
n(r-1)+1}(1-\frac{k+1}{n(r-1)}) \geq 1-\frac{k+2}{n(r-1)}
\end{equation*}\qed
\end{proof}

\medskip

We now provide some lower bounds for the Markov chain $\mathcal{M}_{k}^{2}$.

\begin{lemma}
\label{sk-statement-lem}In the Markov chain $\mathcal{M}_{k}^{2}$, for any $%
r>5$,%
\begin{equation}
s_{k}^{k}\geq 1-\frac{64r}{(r-5)(r-1)}\cdot \frac{n}{(n-k)^{2}}
\label{sk-statement}
\end{equation}
\end{lemma}

\begin{proof}
For $i=k$, the value of $s_{i}^{k}$ in the Markov chain $\mathcal{M}_{k}^{2}$
is%
\begin{equation}
s_{k}^{k}=h_{k}^{k}\cdot s_{k+1}^{k}+(1-h_{k}^{k})\cdot 0=h_{k}^{k}\cdot
s_{k+1}^{k}  \label{sk-eq-1}
\end{equation}%
Therefore Lemma~\ref{hk-statement-lem} implies that%
\begin{equation}
s_{k}^{k}\geq (1-\frac{2}{n(r-1)+1})s_{k+1}^{k}\geq (1-\frac{2}{n(r-1)}%
)s_{k+1}^{k}  \label{sk-eq-1b}
\end{equation}%
and thus%
\begin{equation}
s_{k+1}^{k}-s_{k}^{k}\leq \frac{2}{n(r-1)}s_{k+1}^{k}\leq \frac{2}{n(r-1)}
\label{sk-eq-2}
\end{equation}%
Furthermore, for every $i\in \{k+1,\ldots ,n\}$, the value of $s_{i}^{k}$ in 
$\mathcal{M}_{k}^{2}$ can be computed as follows.%
\begin{equation}
s_{i}^{k}=\alpha _{i}^{k}\cdot s_{i+1}^{k}+\beta _{i}^{k}\cdot
s_{i-1}^{k}+\gamma _{i}^{k}\cdot 1  \label{si-eq-1}
\end{equation}%
where%
\begin{eqnarray}
\alpha _{i}^{k} &=&\frac{r\frac{i(n-i)}{n}}{\Sigma _{i}^{k}}  \notag \\
\beta _{i}^{k} &=&\frac{(i-k)+\frac{i(n-i)}{n}}{\Sigma _{i}^{k}}=\frac{\frac{%
(i-k)n+i(n-i)}{n}}{\Sigma _{i}^{k}}  \label{alpha-beta-gamma-si-eq} \\
\gamma _{i}^{k} &=&\frac{r\frac{i-k}{n}}{\Sigma _{i}^{k}}  \notag
\end{eqnarray}%
and $\Sigma _{i}^{k}=r\frac{i(n-i)}{n}+\frac{(i-k)n+i(n-i)}{n}+r\frac{i-k}{n}
$. Therefore~(\ref{si-eq-1}) implies that%
\begin{equation}
s_{i+1}^{k}-s_{i}^{k}=\frac{\beta _{i}^{k}}{\alpha _{i}^{k}}%
(s_{i}^{k}-s_{i-1}^{k})-\frac{\gamma _{i}^{k}}{\alpha _{i}^{k}}(1-s_{i}^{k})
\label{si-eq-2}
\end{equation}%
Furthermore,~(\ref{alpha-beta-gamma-si-eq}) implies that%
\begin{eqnarray}
\frac{\beta _{i}^{k}}{\alpha _{i}^{k}} &=&\frac{1}{r}\cdot (1+\frac{(i-k)n}{%
i(n-i)})  \label{beta-over-alpha-si-eq-1} \\
\frac{\gamma _{i}^{k}}{\alpha _{i}^{k}} &=&\frac{i-k}{i(n-i)}\geq \frac{i-k}{%
i}\cdot \frac{1}{n}  \label{gamma-over-alpha-si-eq}
\end{eqnarray}%
We now prove that $\frac{\beta _{i}^{k}}{\alpha _{i}^{k}}\leq \frac{5}{r}$,
whenever $i\leq \frac{n+k}{2}$. Suppose first that $k\leq \frac{n}{2}$. Then 
$i\leq \frac{n+k}{2}\leq \frac{n+\frac{n}{2}}{2}$, i.e.~$i\leq \frac{3n}{4}$%
. Thus $\frac{1}{n-i}\leq \frac{4}{n}$, and thus~(\ref%
{beta-over-alpha-si-eq-1}) implies that $\frac{\beta _{i}^{k}}{\alpha
_{i}^{k}}\leq \frac{1}{r}\cdot (1+4)=\frac{5}{r}$. Suppose now that $n\geq k>%
\frac{n}{2}$. Then also $i>\frac{n}{2}$, since $i\geq k+1$, and thus $\frac{n%
}{i}<2$. Furthermore $i-k\leq n-i$, since $i\leq \frac{n+k}{2}$. Therefore $%
\frac{(i-k)n}{i(n-i)}=\frac{i-k}{n-i}\cdot \frac{n}{i}<2$, and thus~(\ref%
{beta-over-alpha-si-eq-1}) implies that $\frac{\beta _{i}^{k}}{\alpha
_{i}^{k}}<\frac{1}{r}\cdot (1+2)=\frac{3}{r}$. Summarizing, for every $k\in
\{1,2,\ldots ,n-1\}$ and every $i\in \{k+1,\ldots ,\frac{n+k}{2}\}$,%
\begin{equation}
\frac{\beta _{i}^{k}}{\alpha _{i}^{k}}\leq \frac{5}{r}
\label{beta-over-alpha-si-eq-2}
\end{equation}%
Therefore it follows by~(\ref{si-eq-2}),~(\ref{gamma-over-alpha-si-eq}), and
(\ref{beta-over-alpha-si-eq-2}) that%
\begin{equation}
s_{i+1}^{k}-s_{i}^{k}\leq \frac{5}{r}(s_{i}^{k}-s_{i-1}^{k})-\frac{i-k}{in}%
(1-s_{i}^{k})  \label{si-eq-3}
\end{equation}%
Thus, in particular%
\begin{eqnarray}
s_{i}^{k}-s_{i-1}^{k} &\leq &\frac{5}{r}(s_{i-1}^{k}-s_{i-2}^{k})  \notag \\
&\leq &(\frac{5}{r})^{i-k-1}(s_{k+1}^{k}-s_{k}^{k})  \label{si-eq-4}
\end{eqnarray}%
Now~(\ref{si-eq-3}) and~(\ref{si-eq-4}) imply that%
\begin{equation}
s_{i+1}^{k}-s_{i}^{k}\leq (\frac{5}{r})^{i-k}(s_{k+1}^{k}-s_{k}^{k})-\frac{%
i-k}{in}(1-s_{i}^{k})  \label{si-eq-5}
\end{equation}%
Note that $%
s_{i}^{k}=s_{k}^{k}+(s_{k+1}^{k}-s_{k}^{k})+(s_{k+2}^{k}-s_{k+1}^{k})+\ldots
+(s_{i}^{k}-s_{i-1}^{k})$. Thus~(\ref{si-eq-4}) implies that%
\begin{eqnarray}
s_{i}^{k} &\leq &s_{k}^{k}+(s_{k+1}^{k}-s_{k}^{k})\cdot (1+\frac{5}{r}%
+\ldots +(\frac{5}{r})^{i-k-1})  \notag \\
&\leq &s_{k}^{k}+(s_{k+1}^{k}-s_{k}^{k})\cdot \frac{r}{r-5}  \label{si-eq-6}
\end{eqnarray}%
Therefore~(\ref{si-eq-5}) and~(\ref{si-eq-6}) imply that%
\begin{equation}
s_{i+1}^{k}-s_{i}^{k}\leq (\frac{5}{r})^{i-k}(s_{k+1}^{k}-s_{k}^{k})-\frac{%
i-k}{in}(1-s_{k}^{k}-(s_{k+1}^{k}-s_{k}^{k})\cdot \frac{r}{r-5})
\label{si-eq-7}
\end{equation}%
Note that~(\ref{si-eq-7}) holds also for $i=k$ and that in this case it
becomes an equality. Summing up~(\ref{si-eq-7}) for every $i\in \{k,\ldots ,%
\frac{n+k}{2}\}$, it follows that%
\begin{equation}
s_{\frac{n+k}{2}+1}^{k}-s_{k}^{k} \leq 
\frac{r}{r-5}(s_{k+1}^{k}-s_{k}^{k})-(1-s_{k}^{k}-(s_{k+1}^{k}-s_{k}^{k})\cdot \frac{r}{%
r-5})\sum_{i=k+1}^{\frac{n+k}{2}}\frac{i-k}{in}  \label{si-eq-8}
\end{equation}%
Note now that for any positive numbers $x,y,z,w>0$, it holds that $\frac{x}{y%
}+\frac{z}{w}>\frac{x+z}{y+w}$. Therefore, for every $i\in \{k+1,\ldots ,%
\frac{n+k}{2}\}$,%
\begin{equation*}
\frac{i-k}{in}+\frac{(\frac{n+k}{2}-i+k+1)-k}{(\frac{n+k}{2}-i+k+1)n} >%
\frac{n-k+2}{n(n+3k+2)}>\frac{n-k}{n(n+3k)}
\end{equation*}%
Thus%
\begin{equation*}
2\sum_{i=k+1}^{\frac{n+k}{2}}\frac{i-k}{in}>(\frac{n+k}{2}-k)\cdot \frac{n-k%
}{n(n+3k)}=\frac{(n-k)^{2}}{2n(n+3k)}
\end{equation*}%
i.e.%
\begin{equation}
\sum_{i=k+1}^{\frac{n+k}{2}}\frac{i-k}{in}>\frac{(n-k)^{2}}{4n(n+3k)}
\label{si-eq-9}
\end{equation}%
It follows now by~(\ref{si-eq-8}) and~(\ref{si-eq-9}) that%
\begin{eqnarray*}
0 \leq s_{\frac{n+k}{2}+1}^{k}-s_{k}^{k}&\leq& \frac{r}{r-5}%
(s_{k+1}^{k}-s_{k}^{k})-\frac{(n-k)^{2}}{4n(n+3k)}%
(1-s_{k}^{k}-(s_{k+1}^{k}-s_{k}^{k})\cdot \frac{r}{r-5}) \\
&=&\frac{r}{r-5}(s_{k+1}^{k}-s_{k}^{k})\frac{4n(n+3k)+(n-k)^{2}}{4n(n+3k)}-%
\frac{(n-k)^{2}}{4n(n+3k)}(1-s_{k}^{k})
\end{eqnarray*}%
Therefore%
\begin{equation*}
(n-k)^{2}(1-s_{k}^{k})\leq \frac{r}{r-5}%
(s_{k+1}^{k}-s_{k}^{k})(4n(n+3k)+(n-k)^{2})
\end{equation*}%
and thus%
\begin{equation}
s_{k}^{k}\geq 1-\frac{r}{r-5}(s_{k+1}^{k}-s_{k}^{k})(1+\frac{4n(n+3k)}{%
(n-k)^{2}})  \label{si-eq-10}
\end{equation}%
Now~(\ref{si-eq-10}) and~(\ref{sk-eq-2}) imply that%
\begin{eqnarray*}
s_{k}^{k} &\geq &1-\frac{r}{r-5}\frac{2}{n(r-1)}(1+\frac{4n(n+3k)}{(n-k)^{2}}%
)\geq 1-\frac{r}{r-5}\frac{2}{n(r-1)}\cdot 2\frac{4n(n+3k)}{(n-k)^{2}} \\
&\geq& 1-\frac{64r}{%
(r-5)(r-1)}\cdot \frac{n}{(n-k)^{2}}
\end{eqnarray*}\qed
\end{proof}

\medskip

The next two corollaries follow now from Lemma~\ref{sk-statement-lem} by
direct substitution.

\begin{corollary}
\label{sk-n-over-3-statement-cor-1}In the Markov chain $\mathcal{M}_{k}^{2}$%
, for any $r>5$ and any $k\leq \frac{n}{2}$,%
\begin{equation*}
s_{k}^{k}\geq 1-\frac{64r}{(r-5)(r-1)}\cdot \frac{4}{n}=1-O(\frac{1}{n})
\end{equation*}
\end{corollary}

\begin{corollary}
\label{sk-n-over-3-statement-cor-2}In the Markov chain $\mathcal{M}_{k}^{2}$%
, for any $r>5$ and any $k\leq n-\sqrt{n\log n}$,%
\begin{equation*}
s_{k}^{k}\geq 1-\frac{64r}{(r-5)(r-1)}\cdot \frac{1}{\log n}=1-O(\frac{1}{%
\log n})
\end{equation*}
\end{corollary}

We now present an auxiliary lemma that provides a lower bound for
the probability $s_{k}^{k}$, for any $k\leq n-1$.

\begin{lemma}
\label{sk-statement-trivial-lem}In the Markov chain $\mathcal{M}_{k}^{2}$,
for any $r>5$ and any $k\leq n-1$,%
\begin{equation*}
s_{k}^{k}\geq \frac{1}{n}
\end{equation*}
\end{lemma}

\begin{proof}
Let $1\leq k\leq n-1$. Recall by~(\ref{si-eq-1}) and~(\ref%
{alpha-beta-gamma-si-eq}) in the proof of Lemma~\ref{sk-statement-lem} that
for $i=k+1$, 
\begin{equation*}
s_{k+1}^{k}=\frac{\alpha _{k+1}^{k}\cdot s_{k+2}^{k}+\beta _{k+1}^{k}\cdot
s_{k}^{k}+\gamma _{k+1}^{k}\cdot 1}{\alpha _{k+1}^{k}+\beta
_{k+1}^{k}+\gamma _{k+1}^{k}}
\end{equation*}%
Therefore, since $s_{k+2}^{k}\geq s_{k+1}^{k}$, it follows that%
\begin{equation}
s_{k+1}^{k}\geq \frac{\beta _{k+1}^{k}\cdot s_{k}^{k}+\gamma _{k+1}^{k}\cdot
1}{\beta _{k+1}^{k}+\gamma _{k+1}^{k}}  \label{si-eq-12}
\end{equation}%
In particular, it follows by~(\ref{si-eq-12}) and~(\ref%
{alpha-beta-gamma-si-eq}) for $i=k+1$ that%
\begin{equation}
s_{k+1}^{k}\geq \frac{(n+(k+1)(n-k-1))\cdot s_{k}^{k}+r\cdot 1}{%
n+(k+1)(n-k-1)+r}  \label{si-eq-13}
\end{equation}%
Furthermore recall by~(\ref{sk-eq-1b}) in the proof of Lemma~\ref%
{sk-statement-lem} that 
\begin{equation}
s_{k+1}^{k}\leq \frac{n(r-1)}{n(r-1)-2}s_{k}^{k}=(1+\frac{2}{n(r-1)-2}%
)s_{k}^{k}  \label{si-eq-14}
\end{equation}%
Thus~(\ref{si-eq-13}) and~(\ref{si-eq-14}) imply that%
\begin{equation*}
(1+\frac{2}{n(r-1)-2})s_{k}^{k}\geq \frac{(n+(k+1)(n-k-1))\cdot
s_{k}^{k}+r\cdot 1}{n+(k+1)(n-k-1)+r}
\end{equation*}%
and thus%
\begin{eqnarray*}
(n+(k+1)(n-k-1)+r)(1+\frac{2}{n(r-1)-2})s_{k}^{k} &\geq
&(n+(k+1)(n-k-1))s_{k}^{k}+r\Leftrightarrow \\
(n+(k+1)(n-k-1))\frac{2}{n(r-1)-2}s_{k}^{k}+r(1+\frac{2}{n(r-1)-2})s_{k}^{k}
&\geq &r\Leftrightarrow \\
(2(n+(k+1)(n-k-1))+rn(r-1))s_{k}^{k} &\geq &r(n(r-1)-2)
\end{eqnarray*}%
Therefore%
\begin{equation*}
s_{k}^{k}\geq \frac{r(n(r-1)-2)}{2(n+(k+1)(n-k-1))+rn(r-1)}
\end{equation*}%
Note now that $(k+1)(n-k-1)<n^{2}$, and thus the last inequality implies that%
\begin{equation*}
s_{k}^{k}\geq \frac{r(n(r-1)-2)}{2(n+n^{2})+rn(r-1)}\geq \frac{r(r-1)-\frac{%
2r}{n}}{2(n+1)+r(r-1)}
\end{equation*}%
Therefore, since $r>5$ and $r<n$ by assumption, it follows that%
\begin{equation*}
s_{k}^{k}\geq \frac{20-2}{2(n+1)+20}=\frac{9}{n+11}>\frac{1}{n}
\end{equation*}\qed
\end{proof}

\subsubsection{Urchin graphs are strong selective amplifiers\label%
{urchin-selective-subsubsec}}

\begin{figure}[tbh]
\centering\includegraphics[scale=0.68]{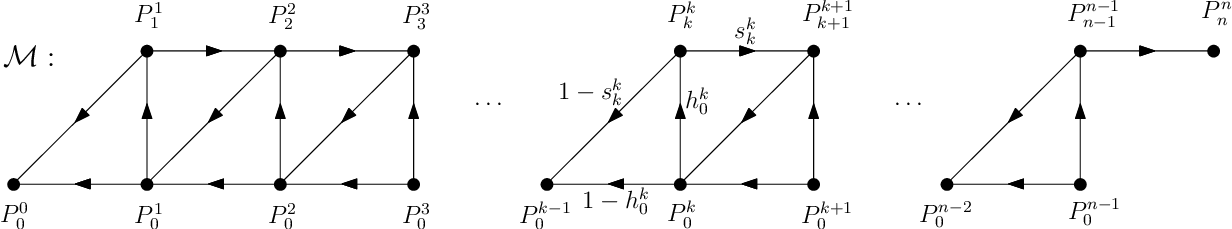}
\caption{The Markov chain $\mathcal{M}$, using the Markov chains $\mathcal{M}%
_{k}^{1}$ and $\mathcal{M}_{k}^{2} $, where $k\in \{1,2,\ldots ,n-1\}$.}
\label{state-graph-M-full-fig}
\end{figure}

In this section we conclude our analysis by combining the results of Section~%
\ref{two-Markov-chains-subsubsec} on the two Markov chains $\mathcal{M}^{1}_{k}$
and $\mathcal{M}^{2}_{k}$. The Markov chain $\mathcal{M}$ is illustrated in
Figure~\ref{state-graph-M-full-fig}, where the transition from state $%
P_{0}^{k}$ to the states $P_{k}^{k},P_{0}^{k-1}$ is done through the Markov
chain $\mathcal{M}^{1}_{k}$, and the transition from state $P_{k}^{k}$ to the
states $P_{k+1}^{k+1},P_{0}^{k-1}$ is done through the Markov chain $\mathcal{M}^{2}_{k}$, respectively.

In Figure~\ref{state-graph-M-full-fig}, the transition probability from
state $P_{k}^{k}$ to state $P_{k+1}^{k+1}$ (resp.~$P_{0}^{k-1}$) is $%
s_{k}^{k}$ (resp.~$1-s_{k}^{k}$). Recall that $s_{k}^{k}$ is the probability
that, starting at $P_{k}^{k}$ in $\mathcal{M}^{2}_{k}$ (and thus also in $%
\mathcal{M}$), we reach state $F_{k+1}$ before we reach $F_{k-1}$.
Furthermore, the transition probability from state $P_{0}^{k}$ to state $%
P_{k}^{k}$ is equal to the probability that, starting at $P_{0}^{k}$ in $\mathcal{M}^{1}_{k}$, we reach $P_{k}^{k}$ before we reach $F_{k-1}$. Note that
this probability is larger than $h_{0}^{k}$. Therefore, in order to compute
a lower bound of the fixation probability of a nose in $G_{n}$, we can
assume that in $\mathcal{M}$ the transition probability from state $%
P_{0}^{k} $ to $P_{k}^{k}$ (resp.~$P_{0}^{k-1}$) is $h_{0}^{k}$ (resp.~$%
1-h_{0}^{k}$), as it is shown in Figure~\ref{state-graph-M-full-fig}.

\begin{figure}[tbh]
\centering%
\subfigure[]{ \label{state-graph-M-prime-full-fig-1}
\includegraphics[scale=0.68]{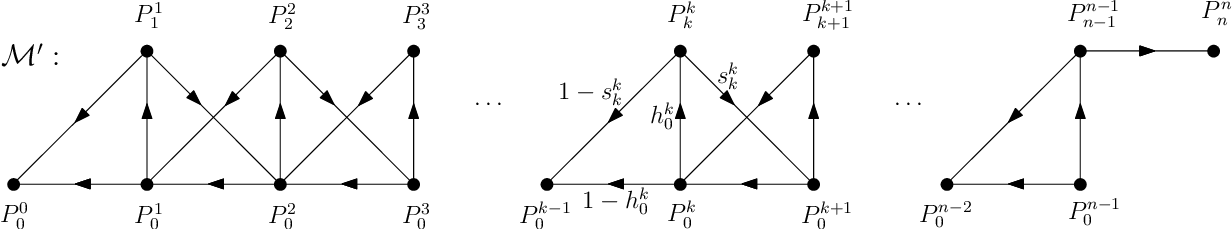}} \hspace{%
0.1cm} 
\subfigure[]{ \label{state-graph-M-prime-full-fig-2}
\includegraphics[scale=0.68]{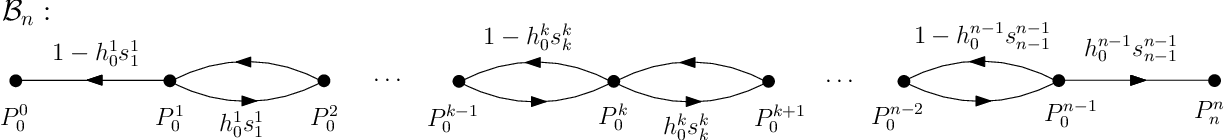}}
\caption{(a) The relaxed Markov chain $\mathcal{M}^{\prime}$ and (b) the
birth-death process $\mathcal{B}_{n}$ that is obtained from $\mathcal{M}%
^{\prime}$ after eliminating the states $P_{k}^{k}$ in $\mathcal{M}^{\prime
} $, where $k\in \{1,2,\ldots,n-1\}$.}
\label{state-graph-M-prime-full-fig}
\end{figure}

Note that for every $k\in \{2,\ldots ,n-1\}$ the infected vertices of state $%
P_{0}^{k}$ is a strict subset of the infected vertices of state $P_{k}^{k}$.
Therefore, in order to compute a lower bound of the fixation probability of
state $P_{0}^{1}$ in $\mathcal{M}$, we can relax $\mathcal{M}$ by changing
every transition from state $P_{k-1}^{k-1}$ to state $P_{k}^{k}$ to a
transition from state $P_{k-1}^{k-1}$ to state $P_{0}^{k}$, where $k\in
\{2,\ldots ,n-1\}$. This relaxed Markov chain $\mathcal{M}^{\prime }$ is
illustrate in Figure~\ref{state-graph-M-prime-full-fig-1}. After eliminating
the states $P_{k}^{k}$ in $\mathcal{M}^{\prime }$, where $k\in \{1,2,\ldots
,n-1\}$, we obtain the equivalent birth-death process $\mathcal{B}_{n}$ that
is illustrated in Figure~\ref{state-graph-M-prime-full-fig-2}. Denote by $%
p_{1}$ the fixation probability of state $P_{0}^{1}$ in $\mathcal{B}_{n}$,
i.e.~$p_{1}$ is the probability that, starting at state $P_{0}^{1}$ in $%
\mathcal{B}_{n}$, we eventually arrive to state $P_{n}^{n}$.

\begin{theorem}
\label{birth-death-lower-bound-thm}For any $r>5$ and for sufficiently large $%
n$, the fixation probability $p_{1}$ of state $P_{0}^{1}$ in $\mathcal{B}%
_{n} $ is $p_{1}\geq 1-\frac{c(r)}{n}$, for some appropriate function $c(r)$
of $r $.
\end{theorem}

\begin{proof}
Denote by $\lambda _{k}$ the forward bias of $\mathcal{B}_{n}$ at state $%
P_{0}^{k}$, i.e.~$\lambda _{k}=\frac{h_{0}^{k}s_{k}^{k}}{1-h_{0}^{k}s_{k}^{k}%
}$ is the ratio of the forward over the backward transition probability at
state $P_{0}^{k}$. Then the fixation probability $p_{1}$ of state $P_{0}^{1}$
in $\mathcal{B}_{n}$ is 
\begin{equation}
p_{1}=\frac{1}{1+\frac{1}{\lambda _{1}}+\frac{1}{\lambda _{1}\lambda _{2}}+%
\frac{1}{\lambda _{1}\lambda _{2}\lambda _{3}}+\ldots +\frac{1}{\lambda
_{1}\lambda _{2}\lambda _{3}\ldots \lambda _{n-1}}}  \label{p-si-eq-2}
\end{equation}%
Note now by Lemma~\ref{h-0-k-statement-lem} and Corollary~\ref%
{sk-n-over-3-statement-cor-1} that for every $k\leq \frac{n}{2}$,%
\begin{eqnarray}
\lambda _{k} &=&\frac{h_{0}^{k}s_{k}^{k}}{1-h_{0}^{k}s_{k}^{k}}\geq \frac{(1-%
\frac{k+2}{n(r-1)})(1-\frac{256r}{(r-5)(r-1)}\cdot \frac{1}{n})}{1-(1-\frac{%
k+2}{n(r-1)})(1-\frac{256r}{(r-5)(r-1)}\cdot \frac{1}{n})}  \notag \\
&\geq &\frac{1-\frac{256r}{(r-5)(r-1)}\cdot \frac{1}{n}-\frac{k+2}{n(r-1)}}{%
\frac{256r}{(r-5)(r-1)}\cdot \frac{1}{n}+\frac{k+2}{n(r-1)}}=\frac{n(r-1)-%
\frac{256r}{(r-5)}-(k+2)}{\frac{256r}{(r-5)}+(k+2)}  \label{lamda-eq0}
\end{eqnarray}%
Therefore, since $k\leq \frac{n}{2}$ and $\frac{256r}{(r-5)}<\log n<\frac{n}{%
2}-2$ for sufficiently large $n$, it follows by~(\ref{lamda-eq0}) that%
\begin{equation*}
\lambda _{k}>\frac{n(r-2)}{\log n+(k+2)}>\frac{n(r-2)}{2\log n+k}
\end{equation*}%
and thus%
\begin{equation}
\frac{1}{\lambda _{k}}<\frac{2\log n+k}{n(r-2)}  \label{lamda-eq1}
\end{equation}%
for every $k\leq \frac{n}{2}$. Furthermore, note by Lemma~\ref%
{h-0-k-statement-lem} and Corollary~\ref{sk-n-over-3-statement-cor-2} that,
whenever $\frac{n}{2}<k\leq n-\sqrt{n\log n}$,%
\begin{eqnarray*}
\lambda _{k} &=&\frac{h_{0}^{k}s_{k}^{k}}{1-h_{0}^{k}s_{k}^{k}}\geq \frac{(1-%
\frac{k+2}{n(r-1)})(1-\frac{64r}{(r-5)(r-1)}\cdot \frac{1}{\log n})}{1-(1-%
\frac{k+2}{n(r-1)})(1-\frac{64r}{(r-5)(r-1)}\cdot \frac{1}{\log n})} \\
&\geq &\frac{1-\frac{64r}{(r-5)(r-1)}\cdot \frac{1}{\log n}-\frac{k+2}{n(r-1)%
}}{\frac{64r}{(r-5)(r-1)}\cdot \frac{1}{\log n}+\frac{k+2}{n(r-1)}}=\frac{%
(r-1)\log n-\frac{64r}{(r-5)}-\frac{k+2}{n}\log n}{\frac{64r}{(r-5)}+\frac{%
k+2}{n}\log n}
\end{eqnarray*}%
Therefore, since $k+2<n$ and $\frac{64r}{(r-5)}<\frac{\log n}{r-2}$ for
sufficiently large $n$, it follows that%
\begin{equation*}
\lambda _{k}>\frac{(r-1)\log n-\frac{\log n}{r-2}-\log n}{\frac{\log n}{r-2}%
+\log n}=r-3
\end{equation*}%
and thus%
\begin{equation}
\frac{1}{\lambda _{k}}\leq \frac{1}{r-3}  \label{lamda-eq2}
\end{equation}%
whenever $\frac{n}{2}<k\leq n-\sqrt{n\log n}$. Moreover, note by Lemma~\ref%
{h-0-k-statement-lem} and Lemma~\ref{sk-statement-trivial-lem} that,
whenever $n-\sqrt{n\log n}<k\leq n-1$,%
\begin{equation*}
\lambda _{k} =\frac{h_{0}^{k}s_{k}^{k}}{1-h_{0}^{k}s_{k}^{k}}\geq \frac{(1-%
\frac{k+2}{n(r-1)})\frac{1}{n}}{1-(1-\frac{k+2}{n(r-1)})\frac{1}{n}} > 
\frac{\frac{1}{n}-\frac{n+1}{n^{2}(r-1)}}{1-\frac{1}{n}+\frac{n+1}{%
n^{2}(r-1)}}>\frac{1-\frac{2}{r-1}}{n-1+\frac{2}{r-1}}
\end{equation*}%
Note now that $1-\frac{2}{r-1}>\frac{1}{2}$ since $r>5$ by assumption, and
thus the latter inequality implies that $\lambda _{k}>\frac{1/2}{n}$, i.e.%
\begin{equation}
\frac{1}{\lambda _{k}}<2n  \label{lamda-eq3}
\end{equation}%
whenever $n-\sqrt{n\log n}<k\leq n-1$.

Since $r>5$ by assumption, note now by~(\ref{lamda-eq1}) that $\frac{1}{%
\lambda _{k}}<\frac{3\log n}{n(r-2)}<\frac{\log n}{n}$ whenever $k\leq \log
n $, and that $\frac{1}{\lambda _{k}}<\frac{3k}{n(r-2)}<\frac{k}{n}$
whenever $\log n<k\leq \frac{n}{2}$. Therefore, for every $k\in \{2,3,\ldots
,\log n\}$,%
\begin{equation}
\frac{1}{\lambda _{1}\lambda _{2}\lambda _{3}\ldots \lambda _{k}}<\left( 
\frac{\log n}{n}\right) ^{k}\leq \left( \frac{\log n}{n}\right) ^{2}
\label{lamda-product-1}
\end{equation}%
Furthermore, for every $k\in \{\log n+1,\ldots ,\frac{n}{2}\}$,%
\begin{eqnarray}
\frac{1}{\lambda _{1}\lambda _{2}\lambda _{3}\ldots \lambda _{k}} &<&\left( 
\frac{\log n}{n}\right) ^{\log n}\prod_{i=\log n+1}^{k}\frac{i}{n}
\label{lamda-product-2-1} \\
&<&\left( \frac{\log n}{n}\right) ^{\log n}<\left( \frac{\log n}{n}\right)
^{3}  \label{lamda-product-2-2}
\end{eqnarray}%
Therefore, for every $k\in \{2,3,\ldots ,\frac{n}{2}\}$,%
\begin{equation}
\sum_{k=2}^{\frac{n}{2}}\frac{1}{\lambda _{1}\lambda _{2}\lambda _{3}\ldots
\lambda _{k}} <\log n\left( \frac{\log n}{n}\right) ^{2}+\frac{n}{2}\left( 
\frac{\log n}{n}\right) ^{3}  <\frac{1}{n}  \label{Sigma-lamda-eq-1}
\end{equation}%
for sufficiently large $n$. Note furthermore by~(\ref{lamda-eq2}) that $%
\frac{1}{\lambda _{k}}<1$ whenever $\frac{n}{2}<k\leq n-\sqrt{n\log n}$,
since $r>5$ by assumption. Therefore, for every $k\in \{\frac{n}{2}+1,\ldots
,n-\sqrt{n\log n}\}$,%
\begin{equation*}
\frac{1}{\lambda _{1}\lambda _{2}\lambda _{3}\ldots \lambda _{k}}<\frac{1}{%
\lambda _{1}\lambda _{2}\lambda _{3}\ldots \lambda _{2n/3}}
\end{equation*}%
and thus it follows by~(\ref{lamda-product-2-2}) that%
\begin{equation}
\sum_{k=\frac{n}{2}+1}^{n-\sqrt{n\log n}}\frac{1}{\lambda _{1}\lambda
_{2}\lambda _{3}\ldots \lambda _{k}}<\left( \frac{n}{2}-\sqrt{n\log n}%
\right) \left( \frac{\log n}{n}\right) ^{3}<\frac{\log ^{3}n}{2n^{2}}<\frac{1%
}{n}  \label{Sigma-lamda-eq-2}
\end{equation}%
for sufficiently large $n$. Let now $n-\sqrt{n\log n}<k\leq n-1$. Then it
follows by~(\ref{lamda-eq3}) and~(\ref{lamda-product-2-1}) that%
\begin{eqnarray}
\frac{1}{\lambda _{1}\lambda _{2}\lambda _{3}\ldots \lambda _{k}} &<&\left( 
\frac{\log n}{n}\right) ^{\log n}\prod_{i=\log n+1}^{\frac{n}{2}}\frac{i}{n}%
\cdot \left( 2n\right) ^{k-n+\sqrt{n\log n}}  \notag \\
&<&\frac{2^{\left( \log n\log \log n+\sqrt{n\log n}+\log n\sqrt{n\log n}%
\right) }}{2^{n/2}}  \label{lamda-product-3}
\end{eqnarray}%
However%
\begin{equation*}
\log n\log \log n+\sqrt{n\log n}+\log n\sqrt{n\log n}<\frac{n}{4}
\end{equation*}%
for sufficiently large $n$, and thus~(\ref{lamda-product-3}) implies that 
\begin{equation*}
\frac{1}{\lambda _{1}\lambda _{2}\lambda _{3}\ldots \lambda _{k}}<\frac{1}{%
2^{n/4}}
\end{equation*}%
for every $k\in \{n-\sqrt{n\log n}+1,\ldots ,n-1\}$. Therefore%
\begin{equation}
\sum_{k=n-\sqrt{n\log n}+1}^{n-1}\frac{1}{\lambda _{1}\lambda _{2}\lambda
_{3}\ldots \lambda _{k}}<\frac{n}{2^{n/4}}<\frac{1}{n}
\label{Sigma-lamda-eq-3}
\end{equation}%
for sufficiently large $n$. Thus, summing up~(\ref{Sigma-lamda-eq-1}),~(\ref%
{Sigma-lamda-eq-2}), and~(\ref{Sigma-lamda-eq-3}), it follows that%
\begin{equation}
\sum_{k=2}^{n-1}\frac{1}{\lambda _{1}\lambda _{2}\lambda _{3}\ldots \lambda
_{k}}<\frac{3}{n}  \label{Sigma-lamda-eq-4}
\end{equation}%
For $k=1$,~(\ref{lamda-eq0}) implies that%
\begin{equation*}
\lambda _{k}\geq \frac{n(r-1)-\frac{256r}{(r-5)}-3}{\frac{256r}{(r-5)}+3}>n\frac{(r-2)(r-5)}{259r-15}
\end{equation*}%
and thus 
\begin{equation}
\frac{1}{\lambda _{k}}<\frac{259r-15}{(r-2)(r-5)}\cdot \frac{1}{n}
\label{Sigma-lamda-eq-5}
\end{equation}

Summarizing, it follows by~(\ref{p-si-eq-2}),~(\ref{Sigma-lamda-eq-4}), and (%
\ref{Sigma-lamda-eq-5}) that%
\begin{equation*}
p_{1}=\frac{1}{1+\frac{1}{n}\cdot \left( 3+\frac{259r-15}{(r-2)(r-5)}\right) 
}\geq 1-\frac{c(r)}{n}
\end{equation*}%
where $c(r) = 3+\frac{259r-15}{(r-2)(r-5)}$ is a function that inly depends on $r$. 
This completes the proof of the theorem.\qed
\end{proof}

\medskip

We are now ready to provide our main result in this section.

\begin{theorem}
\label{urchin-strong-selective-amplifiers-thm}The class $\mathcal{G}%
=\{G_{n}:n\geq 1\}$ of urchin graphs is a class of strong selective
amplifiers.
\end{theorem}

\begin{proof}
Consider the urchin graph $G_{n}$, where $n\geq 1$. Let $v$ be a nose in $%
G_{n}$. Then the fixation probability $f_{r}(v)$ of $v$ in the generalized
Moran process is greater than or equal to the fixation probability of state $%
P_{0}^{1}$ in the Markov chain $\mathcal{M}$ of Figure~\ref%
{state-graph-M-full-fig} (cf.~Corollary~\ref{domination-2-cor} and the
discussion after it in Section~\ref{selective-amplifiers-subsec}).
Furthermore, the fixation probability of state $P_{0}^{1}$ in the Markov
chain $\mathcal{M}$ is greater than or equal to the fixation probability $%
p_{1}$ of state $P_{0}^{1}$ in the birth-death process $\mathcal{B}_{n}$ in
Figure~\ref{state-graph-M-prime-full-fig-2}. Therefore, since $p_{1}\geq 1-%
\frac{c(r)}{n}$ for any $r>5$ by Theorem~\ref{birth-death-lower-bound-thm},
it follows that $f_{r}(v)\geq 1-\frac{c(r)}{n}$ for $r>r_{0}=5$ and
sufficiently large $n$, where $c(r)$ is a function that depends only on $r$.
Finally, since there exist exactly $\frac{n}{2}$ noses in $G_{n}$, it
follows by Definition~\ref{suppressors-def} that the class $\mathcal{G}$ of
urchin graphs is a class of $(\frac{n}{2},n)$-selective amplifiers, and thus 
$\mathcal{G}$ is a class of strong selective amplifiers.\qed
\end{proof}

\section{Suppressor bounds\label{suppressors-sec}}

In this section we prove our lower bound for the fixation
probability of an arbitrary undirected graph, namely the \emph{Thermal
Theorem} (Section~\ref{thermal-subsec}), which generalizes the analysis of
the fixation probability of regular graphs~\cite{Nowak05}. Furthermore we
present for every function $\phi (n)$, where $\phi (n)=\omega (1)$ and $\phi
(n)\leq \sqrt{n}$, a class of $(\frac{n}{\phi(n) + 1},\frac{n}{\phi (n)})$%
-selective suppressors in Section~\ref{selective-suppressor-subsec}.

\subsection{The Thermal Theorem\label{thermal-subsec}}

Consider a graph $G=(V,E)$ and a fitness value $r>1$. Denote by $\mathcal{M}%
_{r}(G)$ the generalized Moran process on $G$ with fitness $r$. Then, for
every subset $S\notin \{\emptyset ,V\}$ of its vertices, the fixation
probability $f_{r}(S)$ of $S$ in~$\mathcal{M}_{r}(G)$ is given by~(\ref%
{generalized-Moran-exact-fixation-eq}), where $f_{r}(\emptyset )=0$ and $%
f_{r}(V)=1$. That is, the fixation probabilities $f_{r}(S)$, where $S\notin
\{\emptyset ,V\}$, are the solution of the linear system~(\ref%
{generalized-Moran-exact-fixation-eq}) with boundary conditions $%
f_{r}(\emptyset )=0$ and $f_{r}(V)=1$.

Suppose that at some iteration of the generalized Moran process the set $S$
of vertices are infected and that the edge $xy\in E$ (where $x\in S$ and $%
y\notin S$) is activated, i.e.~either $x$ infects $y$ or $y$ disinfects $x$.
Then~(\ref{generalized-Moran-exact-fixation-eq}) implies that the
probability that $x$ infects $y$ is higher if $\frac{1}{\deg x}$ is large;
similarly, the probability that $y$ disinfects $x$ is higher if $\frac{1}{%
\deg y}$ is large. Therefore, in a fashion similar to~\cite{Nowak05}, we
call for every vertex $v\in V$ the quantity $\frac{1}{\deg v}$ the \emph{%
temperature} of $v$: a ``hot'' vertex (i.e.~with high temperature) affects
more often its neighbors than a ``cold'' vertex (i.e.~with low temperature).

Before we proceed, recall that $f_{r}(\emptyset )=0$ and $f_{r}(V)=1$. 
Furthermore, recall by (\ref{generalized-Moran-exact-fixation-eq}) that, 
for every vertex subset $S\notin \{\emptyset ,V\}$,
\begin{equation*}
f_{r}(S)=\frac{\sum_{xy\in E,x\in S,y\notin S}\left( \frac{r}{\deg x}%
f_{r}(S+y)+\frac{1}{\deg y}f_{r}(S-x)\right) }{\sum_{xy\in E,x\in S,y\notin
S}\left( \frac{r}{\deg x}+\frac{1}{\deg y}\right)}
\end{equation*}%
Note that the summation in both the nominator and the denominator is done over all edges $xy\in E$ 
which have one endpoint $x\in S$ and one endpoint $y\notin S$. 
Therefore, for every subset $S\notin \{\emptyset ,V\}$, there exists at least one such pair $x(S),y(S)$ 
of vertices, where $x(S)\in S$, $y(S)\notin S$, and $x(S)y(S)\in E$, such that%
\begin{equation}
f_{r}(S)\geq \frac{\frac{r}{\deg x(S)}f_{r}(S+y(S))+\frac{1}{\deg y(S)}%
f_{r}(S-x(S))}{\frac{r}{\deg x(S)}+\frac{1}{\deg y(S)}}
\label{generalized-Moran-lower-bound-fixation-eq}
\end{equation}%
Thus, solving the linear system that is obtained from~(\ref%
{generalized-Moran-lower-bound-fixation-eq}) by replacing inequalities with
equalities, we obtain a lower bound for the fixation probabilities $f_{r}(S)$, 
where $S\notin \{\emptyset ,V\}$.

In the next definition we introduce a weighted generalization of this linear system, 
which is a crucial tool for our analysis in obtaining the Thermal Theorem.
Note that in Definition~\ref{system-L0-def}, as well as in the remainder of this section, 
for every vertex subset $S\notin \{\emptyset ,V\}$ 
we consider the above two vertices $x(S)$ and $y(S)$ as fixed.

\begin{definition}[the linear system $L_{0}$]
\label{system-L0-def}Let $G=(V,E)$ be an undirected graph and $r>1$. Let
every vertex $v\in V$ have weight \emph{(temperature)} $d_{v}>0$. The linear
system $L_{0}$ on the variables $p_{r}(S)$, where $S\subseteq V$, is given
by the following equations whenever $S\notin \{\emptyset ,V\}$:%
\begin{equation}
p_{r}(S)=\frac{rd_{x(S)}p_{r}(S+y(S))+d_{y(S)}p_{r}(S-x(S))}{%
rd_{x(S)}+d_{y(S)}}  \label{system-L0-def-eq}
\end{equation}%
with boundary conditions $p_{r}(\emptyset )=0$ and $p_{r}(V)=1$.
\end{definition}

With a slight abuse of notation, whenever $S=\{u_{1},u_{2},\ldots ,u_{k}\}$,
we denote $p_{r}(u_{1},u_{2},\ldots ,u_{k})=p_{r}(S)$.

\begin{observation}
\label{Markov-chain-M-0-obs}The linear system $L_{0}$ in Definition~\ref%
{system-L0-def} corresponds naturally to the Markov chain $\mathcal{M}_{0}$
with one state for every subset $S\subseteq V$, where the states $\emptyset $
and $V$ are absorbing, and every non-absorbing state $S$ has exactly two
transitions to the states $S+y(S)$ and $S-x(S)$ with transition
probabilities $q_{S}=\frac{rd_{x(S)}}{rd_{x(S)}+d_{y(S)}}$ and $1-q_{S}$,
respectively.
\end{observation}

\begin{observation}
\label{DW-Moran-obs}Let $G=(V,E)$ be a graph and $r>1$. For every vertex $%
x\in V$ let $d_{x}=\frac{1}{\deg x}$ be the temperature of $x$. Then $%
f_{r}(S)\geq p_{r}(S)$ for every $S\subseteq V$, where the values $p_{r}(S)$
are the solution of the linear system $L_{0}$.
\end{observation}

Before we provide the Thermal Theorem (Theorem~\ref{thermal-thm}), we first
prove an auxiliary result in the next lemma which generalizes the Isothermal
Theorem of~\cite{Nowak05} for regular graphs, i.e.~for graphs with the same
number of neighbors for every vertex.

\begin{lemma}
\label{regular-L0-lem}Let $G=(V,E)$ be a graph with $n$ vertices, $r>1$, and 
$d_{u}$ be the same for all vertices $u\in V$. Then for every vertex $u\in V$%
,%
\begin{equation*}
p_{r}(u)=\frac{1-\frac{1}{r}}{1-\frac{1}{r^{n}}}\geq 1-\frac{1}{r}
\end{equation*}
\end{lemma}

\begin{proof}
Since $d_{u}$ is the same for all vertices $u\in V$, it follows by~(\ref%
{system-L0-def-eq}) that for every set $S\notin \{\emptyset ,V\}$, the
forward probability is $q_{S}=\frac{r}{r+1}$ and the backward probability is 
$1-q_{S}=\frac{1}{r+1}$. Therefore, by symmetry, $p_{r}(S)=p_{r}(S^{\prime
}) $ whenever $|S|=|S^{\prime }|$. For every $0\leq k\leq n$ denote by $%
p_{k}=p_{r}(S)$, where $|S|=k$. Note that $p_{0}=0$ and $p_{n}=1$. Then it
follows by~(\ref{system-L0-def-eq}) that, whenever $1\leq k\leq n-1$,%
\begin{equation*}
p_{k+1}-p_{k}=\frac{1}{r}(p_{k}-p_{k-1})=\ldots =\frac{1}{r^{k}}(p_{1}-p_{0})
\end{equation*}%
Therefore, summing up these equations for every $1\leq k\leq n-1$ it follows
that%
\begin{equation*}
p_{n}-p_{1}=(p_{1}-p_{0})(\frac{1}{r}+\frac{1}{r^{2}}+\ldots +\frac{1}{%
r^{n-1}})
\end{equation*}%
and thus, since $p_{0}=0$ and $p_{n}=1$,%
\begin{equation}
p_{1}=\frac{1}{1+\frac{1}{r}+\frac{1}{r^{2}}+\ldots +\frac{1}{r^{n-1}}}\geq 1-\frac{1}{r}  \notag
\end{equation}\qed
\end{proof}

\medskip

We are now ready to provide our main result in this section which provides
a lower bound for the fixation probability on arbitrary
graphs, parameterized by the maximum ratio between two different
temperatures in the graph.

\begin{theorem}[Thermal Theorem]
\label{thermal-thm}Let $G=(V,E)$ be a connected undirected graph and $r>1$.
Then $f_{r}(v)\geq \frac{r-1}{r+\frac{\deg v}{\deg _{\min }}}$ for every $%
v\in V$.
\end{theorem}

\begin{proof}
Let $G$ have $n$ vertices, i.e.~$|V|=n$. Our proof is based on the linear
system $L_{0}$ of Definition~\ref{system-L0-def}. Namely, we consider the
linear system $L_{0}$ with weight $d_{v}=\frac{1}{\deg v}$ for every vertex $%
v\in V$. Note that $d_{\min }=\frac{1}{\deg _{\max }}$ and $d_{\max }=\frac{1%
}{\deg _{\min }}$. Recall that $\mathcal{M}_{0}$ is the Markov chain that
can be defined from the linear system $L_{0}$ (cf.~Observation~\ref%
{Markov-chain-M-0-obs}), and that every state $S\notin \{\emptyset ,V\}$ of $%
\mathcal{M}_{0}$ has exactly two transitions, namely to states $S+y(S)$ and $%
S-x(S)$.

We now define the Markov chain $\mathcal{M}_{0}^{\ast }$ from $\mathcal{M}%
_{0}$ as follows. Consider an arbitrary state $S\subseteq V$ such that $%
1\leq |S|\leq n-2$. Denote $x(S)=u$ and $y(S)=v$ (note that $v\notin S$).
Furthermore denote $x(S+v)=x_{0}$ and $y(S+v)=y_{0}$. Then perform the
following changes to the Markov chain $\mathcal{M}_{0}$:

\medskip

\textbf{Step A.}~add a new dummy state $X_{S}$ to $\mathcal{M}_{0}$,

\medskip

\textbf{Step B.}~replace the transition from $S$ to $S+v$ by a transition
from $S$ to $X_{S}$ (with the same transition probability $q_{S}$),

\medskip

\textbf{Step C.}~add to state $X_{S}$ the transitions to states $S+v+y_{0}$
and $S+v-x_{0}$, with transition probabilities $q_{S+v}$ and $1-q_{S+v}$,
respectively.

\medskip

An example of the application of the above Steps~A,~B,~C is illustrated in
Figure~\ref{dummy-states-fig}. Denote by $\mathcal{M}_{0}^{\ast }$ the
Markov chain obtained after applying these steps to $\mathcal{M}_{0}$ for
every state $S\subseteq V$ with $1\leq |S|\leq n-2$. Furthermore denote by $%
L_{0}^{\ast }$ the linear set that corresponds to $\mathcal{M}_{0}^{\ast }$.
Note that $L_{0}^{\ast }$ has the additional variables $\{p_{r}(X_{S}):1\leq
|S|\leq n-2\}$ that do not exist in $L_{0}$. Moreover, for every state $%
X_{S} $ of $\mathcal{M}_{0}^{\ast }$, note by the construction of $\mathcal{M%
}_{0}^{\ast }$ that $p_{r}(X_{S})=p_{r}(S+y(S))$ in the solution of $%
L_{0}^{\ast }$.

\begin{figure}[tbh]
\centering%
\subfigure[]{ \label{dummy-states-fig-1}
\includegraphics[scale=0.68]{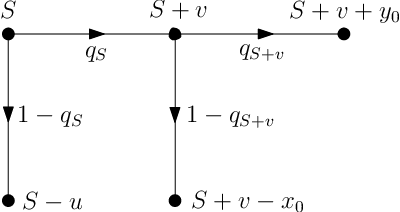}} \hspace{1.0cm} 
\subfigure[]{ \label{dummy-states-fig-2}
\includegraphics[scale=0.68]{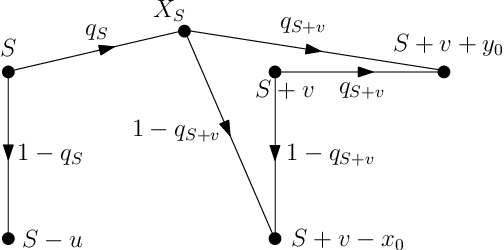}}
\caption{Parts of (a)~the Markov chain $\mathcal{M}_{0}$ and of (b)~the
Markov chain $\mathcal{M}_{0}^{\ast}$ (after the execution of Steps~A,~B,
and~C).}
\label{dummy-states-fig}
\end{figure}

In the remainder of the proof we fix an arbitrary vertex $v_{0}$ and we prove that 
$f_{r}(v_{0})\geq \frac{r-1}{r+\frac{\deg v}{\deg _{\min}}}$. 
To do so, we first consider an arbitrary numbering $v_{0},v_{1},\ldots ,v_{n-1}$ of the $n$
vertices of $G$, starting at our fixed vertex $v_{0}$. 
Then, starting from the Markov chain $\mathcal{M}_{0}^{\ast}$ that we described above 
(cf.~Figure~\ref{dummy-states-fig}), we iteratively construct 
a sequence of Markov chains $\mathcal{M}_{1}^{\ast}, \mathcal{M}_{2}^{\ast}, \ldots, \mathcal{M}_{n-1}^{\ast}$, 
which correspond to a sequence of linear systems $L_{1}^{\ast}, L_{2}^{\ast}, \ldots, L_{n-1}^{\ast}$, 
respectively (cf.~Observation~\ref{Markov-chain-M-0-obs}). 
Each Markov chain~$\mathcal{M}_{i}^{\ast}$ is obtained from the previous chain $\mathcal{M}_{i-1}^{\ast}$ 
by applying two local replacement rules (see Step~1 and Step~2 below) to some transitions 
of $\mathcal{M}_{i-1}^{\ast}$ which involve vertex $v_{i}$. 
Denote by $p_r^{i}(v_0)$ the value of~$p_r(v_0)$ in the solution of the system $L_{i}^{\ast}$. 
We can prove that $p_{r}^{0}(v_{0})\geq \ldots \geq p_{r}^{n-2}(v_{0})\geq p_{r}^{n-1}(v_{0})$. 
Therefore, since $f_{r}(v_{0})\geq p_{r}^{0}(v_{0})$ by Observation~\ref{DW-Moran-obs}, 
each of the values $p_{r}^{i}(v_{0})$ is smaller than or equal to the fixation probability $f_{r}(v_{0})$ of $v$. 
Moreover, it turns out that the value $p_{r}^{i}(v_{0})$ is a \emph{monotone decreasing} function of $d_{v_{i}}$. 
Thus we increase the value of $d_{v_{i}}$ to $d_{\max}=\frac{1}{\deg_{\min}}$ in $L_{i}^{\ast }$ and the value~$p_{r}^{i}(v_{0})$ decreases even more after this change. 
Using this fact we can prove at the end our desired lower bound $f_{r}(v_{0})\geq \frac{r-1}{r+\frac{\deg v}{\deg _{\min}}}$.

For every $i\in \{1,2,\ldots ,n-1\}$, we iteratively construct the Markov chain $\mathcal{M}_{i}^{\ast }$ from the corresponding Markov chain $\mathcal{M}_{i-1}^{\ast }$, as follows. Consider
a state $S\notin \{\emptyset ,V\}$, where $y(S)=v_{i}$ (note that in this
case $v_{i}\notin S$). Denote $x(S)=u$, $x(S+v_{i})=x_{0}$, and $%
y(S+v_{i})=y_{0}$. Then perform the following changes to the Markov chain $%
\mathcal{M}_{i-1}^{\ast }$:

\medskip

\textbf{Step 1.}~replace the transition from $X_{S}$ to $S+v_{i}+y_{0}$ by a
transition from $X_{S}$ to $S+y_{0}$ (with the same transition probability $%
q_{S+v_{i}}$),

\medskip

\textbf{Step 2.}~replace the transition from $X_{S}$ to $S+v_{i}-x_{0}$ by a
transition from $X_{S}$ to $S-x_{0}$ (with the same transition probability $%
1-q_{S+v_{i}}$).

\medskip

Denote by $\mathcal{M}_{i}^{\ast }$ the Markov chain obtained by iteratively
applying Steps~1 and~2 to $\mathcal{M}_{i-1}^{\ast }$ for every state $%
S\notin \{\emptyset ,V\}$ with $y(S)=v_{i}$. Note that for every state $%
S_{1}\notin \{\emptyset ,V\}$ and for every state $S_{2}\neq V$, where $%
v_{i}\in S_{2}$ and $v_{i}\notin S_{1}$, there exists no transition path in $%
\mathcal{M}_{i}^{\ast }$ from $S_{1}$ to $S_{2}$. Furthermore denote by $%
L_{i}^{\ast }$ the linear system that corresponds to $\mathcal{M}_{i}^{\ast
} $ (cf.~Observation~\ref{Markov-chain-M-0-obs}). Note that, in the above
Step~2, if $x_{0}=v_{i}\notin S$ then $S-x_{0}=S$, and thus we have in this
case a transition from $X_{S}$ to $S$ in $\mathcal{M}_{i}^{\ast }$. The
changes made to $\mathcal{M}_{i-1}^{\ast }$ by Steps~1 and~2 are illustrated
in Figures~\ref{step-1-fig} and~\ref{step-2-fig} for the cases where $%
x_{0}=v_{i}$ and $x_{0}\neq v_{i}$, respectively.

\begin{figure}[tbh]
\centering%
\subfigure[]{ \label{step-1-fig-1}
\includegraphics[scale=0.68]{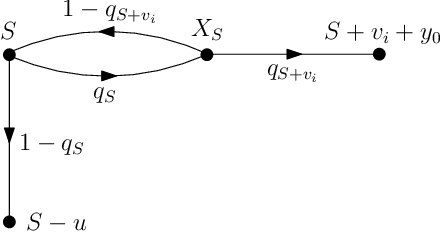}} \hspace{1.0cm} 
\subfigure[]{ \label{step-1-fig-2}
\includegraphics[scale=0.68]{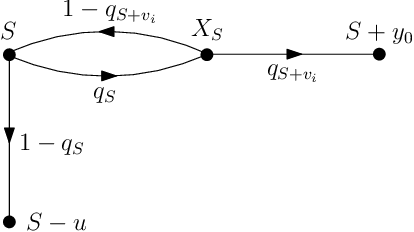}}
\caption{Parts of (a)~the Markov chain $\mathcal{M}_{i-1}^{\ast }$ and of
(b)~the Markov chain $\mathcal{M}_{i}^{\ast }$, where $x_{0}=v_{i}$.}
\label{step-1-fig}
\end{figure}

\begin{figure}[tbh]
\centering%
\subfigure[]{ \label{step-2-fig-1}
\includegraphics[scale=0.68]{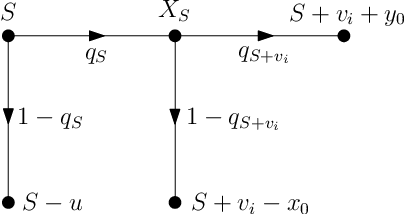}} \hspace{1.0cm} 
\subfigure[]{ \label{step-2-fig-2}
\includegraphics[scale=0.68]{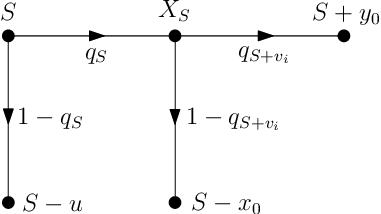}}
\caption{Parts of (a)~the Markov chain $\mathcal{M}_{i-1}^{\ast }$ and of
(b)~the Markov chain $\mathcal{M}_{i}^{\ast }$, where $x_{0}\neq v_{i}$.}
\label{step-2-fig}
\end{figure}

For every state $S$ (resp.~$X_{S}$) of the Markov chain $\mathcal{M}%
_{i}^{\ast }$, and for any $i\in \{0,1,\ldots ,n-1\}$, we denote in the
following by $p_{r}^{i}(S)$ (resp.~$p_{r}^{i}(X_{S})$) the value of $%
p_{r}(S) $ (resp.~$p_{r}(X_{S})$) in the solution of the linear system $%
L_{i}^{\ast }$. Note that for every state $S\notin \{\emptyset ,V\}$ with $%
y(S)=v_{i}$, the transitions of state $X_{S}$ in $\mathcal{M}_{i-1}^{\ast }$
are to the states $S+v_{i}+y_{0}$ and $S+v_{i}-x_{0}$, while the transitions
of state $X_{S}$ in $\mathcal{M}_{i}^{\ast }$ are to the states $%
S+y_{0}\subseteq S+v_{i}+y_{0}$ and $S-x_{0}\subseteq S+v_{i}-x_{0}$,
respectively. Therefore $p_{r}^{i}(X_{S})\leq p_{r}^{i-1}(X_{S})$, and thus $%
p_{r}^{i}(S)\leq p_{r}^{i-1}(S)$. Similarly $p_{r}^{i}(T)\leq p_{r}^{i-1}(T)$%
, for every state $T\subseteq V$ with $v_{i}\notin T$. Thus in particular $%
p_{r}^{i}(v_{j})\leq p_{r}^{i-1}(v_{j})$ for every $j\neq i$.

In order to continue our analysis, we distinguish now the cases where $%
x_{0}=v_{i}$ and $x_{0}\neq v_{i}$.

\medskip

\textbf{Case 1:}~$x_{0}=v_{i}$, cf.~Figure~\ref{step-1-fig}. In this case we
have in the linear system $L_{i}^{\ast }$ that%
\begin{eqnarray*}
p_{r}^{i}(S) &=&q_{S}p_{r}^{i}(X_{S})+(1-q_{S})p_{r}^{i}(S-u) \\
p_{r}^{i}(X_{S}) &=&q_{S+v_{i}}p_{r}^{i}(S+y_{0})+(1-q_{S+v_{i}})p_{r}^{i}(S)
\end{eqnarray*}%
and thus%
\begin{equation*}
p_{r}^{i}(S)=\frac{q_{S}q_{S+v_{i}}p_{r}^{i}(S+y_{0})+(1-q_{S})p_{r}^{i}(S-u)%
}{q_{S}q_{S+v_{i}}+(1-q_{S})}
\end{equation*}%
where%
\begin{eqnarray*}
q_{S} &=&\frac{rd_{u}}{rd_{u}+d_{v_{i}}} \\
q_{S+v_{i}} &=&\frac{rd_{v_{i}}}{rd_{v_{i}}+d_{y_{0}}}
\end{eqnarray*}%
Therefore the forward probability of state $S$ in $\mathcal{M}_{i}^{\ast }$
is (after eliminating the state $X_{S}$) equal to%
\begin{equation}
\frac{q_{S}q_{S+v_{i}}}{q_{S}q_{S+v_{i}}+(1-q_{S})}=\frac{r^{2}d_{u}}{%
r^{2}d_{u}+rd_{v_{i}}+d_{y_{0}}}  \label{forward-case-1-eq}
\end{equation}

\medskip

\textbf{Case 2:}~$x_{0}\neq v_{i}$, cf.~Figure~\ref{step-2-fig}. In this
case we have in the linear system $L_{i}^{\ast }$ that%
\begin{eqnarray*}
p_{r}^{i}(S) &=&q_{S}p_{r}^{i}(X_{S})+(1-q_{S})p_{r}^{i}(S-u) \\
p_{r}^{i}(X_{S})
&=&q_{S+v_{i}}p_{r}^{i}(S+y_{0})+(1-q_{S+v_{i}})p_{r}^{i}(S-x_{0})
\end{eqnarray*}%
and thus%
\begin{equation*}
p_{r}^{i}(S)=q_{S}q_{S+v_{i}}p_{r}^{i}(S+y_{0})+q_{S}(1-q_{S+v_{i}})p_{r}^{i}(S-x_{0})+(1-q_{S})p_{r}^{i}(S-u)
\end{equation*}%
where%
\begin{eqnarray*}
q_{S} &=&\frac{rd_{u}}{rd_{u}+d_{v_{i}}} \\
q_{S+v_{i}} &=&\frac{rd_{x_{0}}}{rd_{x_{0}}+d_{y_{0}}}
\end{eqnarray*}%
Therefore the forward probability of state $S$ in $\mathcal{M}_{i}^{\ast }$
is (after eliminating the state $X_{S}$) equal to%
\begin{equation}
q_{S}q_{S+v_{i}}=\frac{rd_{u}}{rd_{u}+d_{v_{i}}}\cdot \frac{rd_{x_{0}}}{%
rd_{x_{0}}+d_{y_{0}}}  \label{forward-case-2-eq}
\end{equation}

\medskip

It follows now by Cases~1 and~2 (cf.~(\ref{forward-case-1-eq}) and~(\ref%
{forward-case-2-eq})) that the forward probability of state $S$ in $\mathcal{%
M}_{i}^{\ast }$ (after eliminating the state $X_{S}$) is a monotone
decreasing function of $d_{v_{i}}$. Therefore, for every state $S^{\prime
}\subseteq V$ with $v_{i}\notin S^{\prime }$, the value $p_{r}^{i}(S^{\prime
})$ is also a monotone decreasing function of $d_{v_{i}}$. Thus, in
particular, also the value $p_{r}^{i}(v_{j})$, where $j\neq i$, is a
monotone decreasing function of $d_{v_{i}}$. We now increase the value of $%
d_{v_{i}}$ to $d_{\max }$ in $L_{i}^{\ast }$. Thus for every $j\neq i$, the
value $p_{r}^{i}(v_{j})$ decreases after this change.

Recall that $f_{r}(v_{0})\geq p_{r}^{0}(v_{0})$ by Observation~\ref%
{DW-Moran-obs}. Therefore, since also $p_{r}^{0}(v_{0})\geq \ldots \geq
p_{r}^{n-2}(v_{0})\geq p_{r}^{n-1}(v_{0})$, it follows that $%
f_{r}(v_{0})\geq p_{r}^{n-1}(v_{0})$, i.e.~$p_{r}^{n-1}(v_{0})$ is a lower
bound for the fixation probability $f_{r}(v_{0})$ in the Markov chain $%
\mathcal{M}_{r}(G)$. Furthermore $d_{v_{1}}=d_{v_{2}}=\ldots
=d_{v_{n-1}}=d_{\max }$ in the linear system $L_{n-1}^{\ast }$. Consider now
the state $S=\{v_{0}\}$ in the Markov chain $\mathcal{M}_{n-1}^{\ast }$, and
let $y(S)=v_{i_{0}}$, where $1\leq i_{0}\leq n-1$. Note that $x(S)=v_{0}$.
Then the value $p_{r}^{n-1}(v_{0})$ equals%
\begin{eqnarray}
p_{r}^{n-1}(v_{0}) &=&\frac{%
rd_{v_{0}}p_{r}^{n-1}(v_{0},v_{i_{0}})+d_{v_{i_{0}}}p_{r}^{n-1}(\emptyset )}{%
rd_{v_{0}}+d_{v_{i_{0}}}}  \notag \\
&\geq &\frac{rd_{v_{0}}}{rd_{v_{0}}+d_{\max }}p_{r}^{n-1}(v_{i_{0}})  \label{p(v0)-n-1-eq}
\end{eqnarray}%
cf.~Definition~\ref{system-L0-def}. Recall that $d_{v_{0}}=\frac{1}{\deg
v_{0}}$ and $d_{\max }=\frac{1}{\deg _{\min }}$ by definition. Thus, since $%
f_{r}(v_{0})\geq p_{r}^{n-1}(v_{0})$ as we proved above,~(\ref{p(v0)-n-1-eq}%
) implies that%
\begin{equation}
f_{r}(v_{0})\geq p_{r}^{n-1}(v_{0})\geq \frac{r}{r+\frac{\deg v_{0}}{\deg
_{\min }}}p_{r}^{n-1}(v_{i_{0}})  \label{f(v0)-eq}
\end{equation}%
Now, similarly to the above transformations of the linear system $%
L_{i-1}^{\ast }$ to $L_{i}^{\ast }$, where $1\leq i\leq n-1$, we construct
the linear system $L_{n}^{\ast }$ (and the corresponding Markov chain $%
\mathcal{M}_{n}^{\ast }$) from $L_{n-1}^{\ast }$ (and from the corresponding
Markov chain $\mathcal{M}_{n-1}^{\ast }$), by applying iteratively the above
Steps~1 and~2 to the states $S\subseteq V$, where $y(S)=v_{0}$ (instead of $%
y(S)=v_{i}$ above). Furthermore we increase the value of $d_{v_{0}}$ to $%
d_{\max }$ in the resulting linear system $L_{n}^{\ast }$. Then, similarly
to the construction of $L_{i}^{\ast }$, where $1\leq i\leq n-1$, it follows
that $p_{r}^{n}(v_{j})\leq p_{r}^{n-1}(v_{j})$ for every $j\neq 0$. Thus, in
particular, $p_{r}^{n}(v_{i_{0}})\leq p_{r}^{n-1}(v_{i_{0}})$. Furthermore $%
d_{v_{0}}=d_{v_{1}}=\ldots =d_{v_{n-1}}=d_{\max }$ in $L_{n}^{\ast }$, and
thus $p_{r}^{n}(v_{i_{0}})\geq 1-\frac{1}{r}$ by Lemma~\ref{regular-L0-lem}.
Therefore, since $p_{r}^{n}(v_{i_{0}})\leq p_{r}^{n-1}(v_{i_{0}})$, it
follows by~(\ref{f(v0)-eq}) that%
\begin{equation}
f_{r}(v_{0})\geq \frac{r}{r+\frac{\deg v_{0}}{\deg _{\min }}}%
p_{r}^{n}(v_{i_{0}})\geq \frac{(r-1)}{r+\frac{\deg v_{0}}{\deg _{\min }}}
\label{p(v0)-n-1-eq-2}
\end{equation}%
Since $v_{0}$ has been chosen arbitrarily, this completes the proof of the
theorem.\qed
\end{proof}

\medskip

The lower bound for the fixation probability in Theorem~\ref{thermal-thm} is
almost tight. Indeed, if a graph ${G=(V,E)}$ with $n$ vertices is regular,
i.e.~if ${\deg u=\deg v}$ for every ${u,v\in V}$, then ${f_{r}(G)=\frac{1-%
\frac{1}{r}}{1-\frac{1}{r^{n}}}}$ by Lemma~\ref{regular-L0-lem} (cf.~also
the Isothermal Theorem in~\cite{Nowak05}), and thus $f_{r}(G)\cong {\frac{r-1%
}{r}}$ for large enough~$n$. On the other hand, Theorem~\ref{thermal-thm}
implies for a regular graph $G$ that ${f_{r}(G)\geq \frac{r-1}{r+1}}$.

\subsection{A class of selective suppressors\label%
{selective-suppressor-subsec}}

\begin{figure}[h!tb]
\centering 
\subfigure[]{ \label{suppressors-G-n-phi_n-fig}
\includegraphics[scale=0.6]{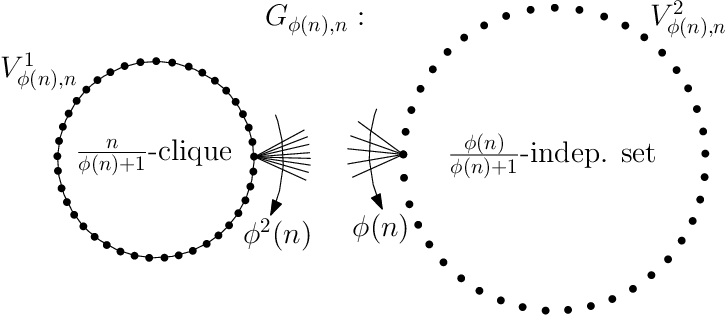}} 
\subfigure[]{ \label{state-graph-clique-fig}
\includegraphics[scale=0.8]{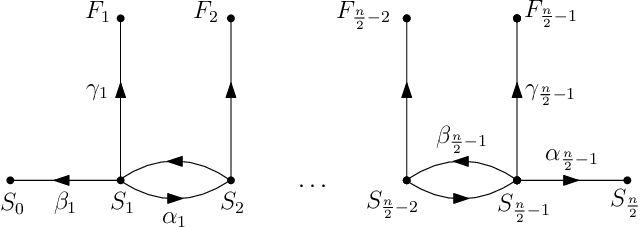}}
\caption{(a) The graph $G_{\protect\phi(n),n}$ with $n$ vertices and (b) the
relaxed Markov chain.}
\label{suppressors-fig}
\end{figure}

In this section we present for every function $\phi (n)$, where $\phi
(n)=\omega (1)$ and $\phi (n)\leq \sqrt{n}$, the class $\mathcal{G}_{\phi
(n)}=\{G_{\phi (n),n}:n\geq 1\}$ of $(\frac{n}{\phi (n)+1},\frac{n}{\phi (n)}%
)$-selective suppressors. 
We call these graphs $\phi(n)$\emph{-urchin graphs}, since for $\phi(n)=1$ they coincide with the class of urchin graphs in Section~\ref{selective-amplifiers-subsec}. 
For every $n$, the graph $G_{\phi (n),n}=(V_{\phi
(n),n},E_{\phi (n),n})$ has $n$ vertices. Its vertex set $V_{\phi (n),n}$
can be partitioned into two sets $V_{\phi (n),n}^{1}$ and $V_{\phi (n),n}^{2}
$, where $|V_{\phi (n),n}^{1}|=\frac{n}{\phi (n)+1}$ and $|V_{\phi
(n),n}^{2}|=\frac{\phi (n)}{\phi (n)+1}n$, such that $V_{\phi (n),n}^{1}$
induces a clique and $V_{\phi (n),n}^{2}$ induces an independent set in $%
G_{\phi (n),n}$. Furthermore, every vertex $u\in V_{\phi (n),n}^{2}$ has $%
\phi (n)$ neighbors in $V_{\phi (n),n}^{1}$, and every vertex $v\in V_{\phi
(n),n}^{1}$ has $\phi ^{2}(n)$ neighbors in $V_{\phi (n),n}^{2}$. Therefore $%
\deg v=n+\phi ^{2}(n)-1$ for every $v\in V_{\phi (n),n}^{1}$ and $\deg
u=\phi (n)$ for every $u\in V_{\phi (n),n}^{2}$. An example of a graph $%
G_{\phi (n),n}$ is illustrated in Figure~\ref{suppressors-G-n-phi_n-fig}.

\begin{lemma}
\label{fixation-clique-upper-bound-lem}For every $v\in V_{\phi (n),n}^{1}$
and sufficiently large $n$,%
\begin{equation*}
f_{r}(v)<5r\cdot \frac{\phi (n)}{n}
\end{equation*}
\end{lemma}

\begin{proof}
Denote by $S_{k}$ the state, in which exactly $k\geq 0$ vertices of $V_{\phi
(n),n}^{1}$ are infected and all vertices of $V_{\phi (n),n}^{2}$ are not
infected. Note that $S_{0}$ is the empty state. Furthermore denote by $F_{k}$
the state where exactly $k\geq 0$ vertices of $V_{\phi (n),n}^{1}$ and at
least one vertex of $V_{\phi (n),n}^{2}$ are infected. In order to compute
an upper bound for the fixation probability $f_{r}(S_{1})$ (i.e.~of the
fixation probability $f_{r}(v)$ where $v\in V_{\phi (n),n}^{1}$), we can set
the value $f_{r}(S_{\frac{n}{2}})$ and the values $f_{r}(F_{k})$ for every $%
k\geq 1$ to their trivial upper bound $1$. That is, we assume that the state 
$S_{\frac{n}{{2}}}$, as well as all states $F_{k}$, where $k\geq 1$, are
absorbing. After performing these relaxations (and eliminating self loops),
we obtain a Markov chain, whose state graph is illustrated in Figure~\ref%
{state-graph-clique-fig}. For any $1\leq k\leq \frac{n}{2}-1$ in this
relaxed Markov chain,%
\begin{equation}
f_{r}(S_{k})=\alpha _{k}f_{r}(S_{k+1})+\beta _{k}f_{r}(S_{k-1})+\gamma _{k}
\label{fixation-clique-eq}
\end{equation}%
where%
\begin{eqnarray}
\alpha _{k} &=&\frac{rk\frac{n-k}{n+\phi ^{2}(n)-1}}{\sum_{k}}  \notag \\
\beta _{k} &=&\frac{k\left( \frac{\phi ^{2}(n)}{\phi (n)}+\frac{n-k}{n+\phi
^{2}(n)-1}\right) }{\sum_{k}}  \label{alpha-beta-gamma} \\
\gamma _{k} &=&\frac{rk\frac{\phi ^{2}(n)}{n+\phi ^{2}(n)-1}}{\sum_{k}} 
\notag
\end{eqnarray}%
where $\sum_{k}=rk\frac{n-k}{n+\phi ^{2}(n)-1}+k\left( \frac{\phi ^{2}(n)}{%
\phi (n)}+\frac{n-k}{n+\phi ^{2}(n)-1}\right) +rk\frac{\phi ^{2}(n)}{n+\phi
^{2}(n)-1}$. Note now by~(\ref{alpha-beta-gamma}) that%
\begin{equation}
\frac{\beta _{k}}{\alpha _{k}}=\frac{1}{r}\left( 1+\frac{\phi ^{2}(n)\left(
n+\phi ^{2}(n)-1\right) }{\phi (n)\left( n-k\right) }\right) >\frac{\phi (n)%
}{r}>1  \label{beta-alpha}
\end{equation}%
Furthermore, since $k<\frac{n}{2}$, it follows by~(\ref{alpha-beta-gamma})
that%
\begin{equation}
\frac{\gamma _{k}}{\alpha _{k}}=\frac{\phi ^{2}(n)}{n-k}<2\frac{\phi ^{2}(n)%
}{n}  \label{gamma-alpha}
\end{equation}%
Now, since $\alpha _{k}+\beta _{k}+\gamma _{k}=1$ and $f_{r}(S_{k})\geq
f_{r}(S_{k-1})$ for every $k$,~(\ref{fixation-clique-eq}) implies by~(\ref%
{beta-alpha}) and~(\ref{gamma-alpha}) that%
\begin{eqnarray*}
f_{r}(S_{k+1})-f_{r}(S_{k}) &=&\frac{\beta _{k}}{\alpha _{k}}%
(f_{r}(S_{k})-f_{r}(S_{k-1}))-\frac{\gamma _{k}}{\alpha _{k}}(1-f_{r}(S_{k}))
\\
&>&\frac{\phi (n)}{r}(f_{r}(S_{k})-f_{r}(S_{k-1}))-2\frac{\phi ^{2}(n)}{n} \\
&>&\ldots  \\
&>&\left( \frac{\phi (n)}{r}\right) ^{k}\cdot f_{r}(S_{1})-2\frac{\phi
^{2}(n)}{n}\cdot \frac{\left( \frac{\phi (n)}{r}\right) ^{k}-1}{\frac{\phi
(n)}{r}-1}
\end{eqnarray*}%
Thus, since $f_{r}(S_{\frac{n}{2}})=1$ in the relaxed Markov chain, we have
that%
\begin{eqnarray*}
1-f_{r}(S_{1}) &=&\sum_{k=1}^{\frac{n}{2}-1}(f_{r}(S_{k+1})-f_{r}(S_{k})) \\
&>&\sum_{k=1}^{\frac{n}{2}-1}\left[ \left( \frac{\phi (n)}{r}\right)
^{k}\cdot f_{r}(S_{1})-2\frac{\phi ^{2}(n)}{n}\cdot \frac{\left( \frac{\phi
(n)}{r}\right) ^{k}-1}{\frac{\phi (n)}{r}-1}\right] 
\end{eqnarray*}%
Therefore%
\begin{equation*}
f_{r}(S_{1})\sum_{k=0}^{\frac{n}{2}-1}\left( \frac{\phi (n)}{r}\right)
^{k}<1+2\frac{\phi ^{2}(n)}{n\left( \frac{\phi (n)}{r}-1\right) }\sum_{k=0}^{%
\frac{n}{2}-1}\left[ \left( \frac{\phi (n)}{r}\right) ^{k}-1\right] 
\end{equation*}%
and thus%
\begin{eqnarray*}
f_{r}(S_{1}) &<&2\frac{\phi ^{2}(n)}{n\left( \frac{\phi (n)}{r}-1\right) }+%
\frac{1}{\sum_{k=0}^{n/2-1}\left( \frac{\phi (n)}{r}\right) ^{k}} \\
&=&2r\frac{\phi ^{2}(n)}{n\left( \phi (n)-r\right) }+\frac{1}{%
\sum_{k=0}^{n/2-1}\left( \frac{\phi (n)}{2}\right) ^{k}}
\end{eqnarray*}%
Therefore, since $\phi (n)=\omega (1)$ and $r$ is constant by assumption, it
follows that $r\leq \frac{\phi (n)}{2}$ for sufficiently large~$n$, and thus%
\begin{equation*}
f_{r}(S_{1})<4r\frac{\phi (n)}{n}+\frac{1}{n}<5r\frac{\phi (n)}{n}
\end{equation*}\qed
\end{proof}

\medskip

Using Lemma~\ref{fixation-clique-upper-bound-lem} we can now prove the next
theorem.

\begin{theorem}
\label{selective-suppressors-thm}For every function $\phi (n)$, where $\phi
(n)=\omega (1)$ and $\phi (n)\leq \sqrt{n}$, the class $\mathcal{G}_{\phi
(n)}=\{G_{\phi (n),n}:n\geq 1\}$ of $\phi(n)$-urchin graphs is a class 
of~$(\frac{n}{\phi (n)+1},\frac{n}{\phi (n)})$-selective suppressors.
\end{theorem}

\begin{proof}
It follows by Lemma~\ref{fixation-clique-upper-bound-lem} that, if $v\in
V_{\phi (n),n}^{1}$, then $f_{r}(v)<5r\frac{\phi (n)}{n}=\frac{5r}{n/\phi (n)%
}$ for any $r>1$ and sufficiently large $n$. Therefore, since $|V_{\phi
(n),n}^{1}|=\frac{n}{\phi (n)+1}$ for every graph $G_{\phi (n),n}$, it
follows by Definition~\ref{suppressors-def} that the class $\mathcal{G}%
_{\phi (n)}$ of graphs is a class of $(\frac{n}{\phi (n)+1},\frac{n}{\phi (n)%
})$-selective suppressors.\qed
\end{proof}

\end{document}